\documentclass{aa}

\usepackage[varg]{txfonts}
\usepackage{natbib}
\bibpunct{(}{)}{;}{a}{}{,} 

\usepackage{hyperref}
\usepackage{graphicx}
\usepackage{booktabs}
\usepackage{multirow}
\usepackage{xcolor}
\usepackage{rotating}
\usepackage[table]{xcolor}
\usepackage{afterpage}
\usepackage[separate-uncertainty=true, multi-part-units=single]{siunitx}
\usepackage{threeparttable}
\usepackage{placeins}

\makeatletter
\renewcommand*\aa@pageof{, page \thepage{} of \pageref*{LastPage}}
\makeatother

\begin{document}

\title{DUETS: Red giant asteroseismic binaries}
\subtitle{Binary confirmation and prospects for double-seismic model inference}

\author{
     J. S. Thomsen\inst{1}\corrauth{j.s.thomsen@pm.me}
    \and W. E. van Rossem\inst{2,1}
    \and A. Miglio\inst{1,3}
    \and A. Mazzi\inst{1}
    \and V. D'Orazi\inst{4,5}
    \and J. Y. Choi\inst{6,7}
    \and W. H. Ball\inst{8}
    \and M. Tailo\inst{1,5}
    \and M. Matteuzzi\inst{9,1,3}
    \and E. Corsaro\inst{10}
    \and K. Brogaard\inst{11}
    \and M. Marcussen\inst{11}
    \and M. N. Lund\inst{11}
    \and F. Grundahl\inst{11}
    \and E. Knudstrup\inst{11}
    \and D. Slumstrup\inst{12,13}
}

\institute{
    Department of Physics and Astronomy "Augusto Righi", University of Bologna, via Gobetti 93/2, 40129, Bologna, Italy
    \and Departament d’Astronomia i Astrofísica, Universitat de València, C. Dr. Moliner 50, 46100 Burjassot, Spain
    \and Osservatorio di Astrofisica e Scienza dello Spazio di Bologna, INAF, Via Gobetti 93/3, Bologna, 40129, Italia
    \and Dept. of Physics, University of Rome Tor Vergata, via della Ricerca Scientifica 1, 00133, Rome, Italy 
    \and INAF Osservatorio Astronomico di Roma, via Frascati 33, 00078, Monte Porzio Catone RM, Italy
    \and Heidelberg Institute for Theoretical Studies (HITS), Schloss-Wolfsbrunnenweg 35, D-69118 Heidelberg, Germany
    \and Heidelberg University, Centre for Astronomy, Landessternwarte, Königstuhl 12, D-69117 Heidelberg, Germany
    \and Advanced Research Computing, University of Birmingham, United Kingdom, B15 2TT
    \and STAR Institute, Université de Liège, Liège, Belgium
    \and INAF – Osservatorio Astrofisico di Catania, Via S. Sofia, 78, 95123 Catania, Italy
    \and Department of Physics and Astronomy, Aarhus University, Ny Munkegade 120, DK-8000 Aarhus C, Denmark
    \and GRANTECAN, Cuesta de San José s/n, E-38712 Breña Baja, La Palma, Spain
    \and Instituto de Astrofísica de Canarias, E-38205 La Laguna, Tenerife, Spain
}

\titlerunning{Asteroseismic binaries confirmed with spectroscopy}

\authorrunning{Thomsen et al.}

\date{}
\abstract{Red giants displaying solar-like oscillations allow precise model-dependent stellar properties to be derived. Asteroseismic binaries (ABs), which are gravitationally bound systems where both stars show detectable oscillations, can provide independent constraints on the interior of two coeval stars, allowing us to model them jointly and potentially obtain constraints unavailable from either star individually. This makes ABs potential laboratories for calibrating uncertain aspects of stellar evolution theory. In this context, AB candidates (ABCs) are systems showing two oscillators, which are as of yet not confirmed as gravitationally bound. The relatively coarse spatial resolution of the \emph{Kepler} mission can blend unrelated oscillators into the same light curve, and past studies have found a high incidence of chance alignments in ABCs due to this. We aim to confirm ABCs and other red giant binary candidates, as well as demonstrate the advantages of AB modelling inference in constraining stellar parameters.
We obtained high-resolution spectroscopic indicators of binarity for a sample of 60 oscillating red giants that have photometric (33 stars) and/or asteroseismic (38) indications of binarity. 
We confirm 12 of our 60 targets as spectroscopic binaries, 8 of which are ABs as well. Cross-matching with Gaia DR3 and APOGEE DR17, we identify a further 9 likely binaries, at least 6 of which are also ABCs. We then investigated the impact of simultaneously modelling the individual oscillation modes of both oscillators of an AB. We inferred stellar properties for: one binary of extragalactic origin containing one detectable red-giant-branch (RGB) oscillator; one AB with RGB oscillators; and two ABCs with RGB oscillators. We find that the mode frequencies of the two ABCs and the confirmed AB are well described with a coeval modelling of the stellar components. We demonstrate that the simultaneous modelling of ABs with two RGB stars can be used to infer the stellar helium content down to 3\% precision, compared to $5\%$ when using only the brightest star. Finally, the metal-poor binary KIC\,2971380 hints at a need to correct the $\nu_{\rm max}$ scaling relation, supporting recent studies of extremely metal-poor oscillators.}


\maketitle
\nolinenumbers
\section{Introduction}
Asteroseismology of solar-like oscillators, including red giants (RGs), allows for highly precise inference of fundamental stellar parameters \citep[e.g. ][]{Chaplin2013}. With photometric monitoring provided by the CoRot, \emph{Kepler}, and TESS space missions \citep{Auvergne2009, Borucki2010, Ricker2014}, thousands of RGs can be studied asteroseismically, allowing us to improve our understanding of both stellar and Galactic evolution \citep{Noels-Grotsch2025}. Particularly for RGs with long-baseline photometric monitoring as provided by the \emph{Kepler} mission, model inferences using individual mode frequencies can be used for Galactic archaeology, as the stellar mass and age can be inferred with better than 2 and 10\% precision, respectively \citep[see e.g.][]{Montalban2021, Thomsen2025}. However, such inferences inherently rely on the use of stellar evolution models. In order to improve the accuracy of these predictions, calibration laboratories are needed to constrain current uncertainties in stellar models \citep[for example, initial helium mass fraction and convection, see e.g.][]{Lebreton2014, Joyce2023}. At the same time, validation benchmarks are needed for testing the predictions of state-of-the-art calibrated models. Binary stars containing a solar-like oscillator, particularly detached eclipsing binaries (dEBs), have been shown to be very suitable for the task of independent validation of the accuracy of asteroseismic predictions \citep[e.g. ][]{Gaulme2016, Brogaard2018, Brogaard2022, Themessl2018, Benbakoura2021, Thomsen2022, Thomsen2025}. This is because dEBs allow the fundamental measurement of stellar masses and radii at a precision down to at least $0.2\%$ \citep{Maxted2020}, without dependence on stellar or binary evolution models \citep[for a review, see][]{Torres2010}. However, the sample size of suitable dEBs is currently small, with fewer than two dozen having \emph{Kepler} observations \citep{Gaulme2016, GaulmeGuzik2019, Benbakoura2021, Hedlund2026}. In the \emph{Kepler} field, a few suitable open clusters also exist that host multiple oscillating RGs with detectable oscillations. Such stellar clusters are each a large laboratory at a fixed age and metallicity, where oscillators and models thereof can both be compared, or benchmarked against independent cluster constraints from dEBs and photometry \citep[see e.g.][]{Brogaard2012,Brogaard2021,Brogaard2023,Handberg2017, Mckeever2019,dufresne2026absoluteageopencluster}. However, their exploration of the mass-metallicity parameter space is limited, and open and globular cluster constraints outside of the \emph{Kepler} field are hampered by lower seismic quality \citep[e.g.][]{Miglio2016,Lund2016,Tailo2022,Howell2024,Howell2026,Stello2026}.

We therefore need more stellar systems that can provide either strong constraints for seismic calibration of stellar models, or independent benchmarking of asteroseismic predictions.
Asteroseismic binaries (ABs, \citet{Miglio2014}), which are gravitationally bound stellar systems where both stars show detectable oscillations, are a unique type of system that can fill the gap for calibrators. Asteroseismic binaries do not necessarily offer stellar-model-independent constraints on fundamental parameters. However, when the two oscillators' mode frequencies can be disentangled, ABs can allow us to perform analysis of two coeval oscillators, typically at different luminosities and/or different evolutionary states, similar to seismology in clusters. Expectations from population synthesis put their numbers in the \emph{Kepler} field at 20-80 ABs following the age and metallicity distribution of the field stars \citep{Mazzi2025}. Most of these are expected to contain two RGs. Uniquely, \citet{Mazzi2025} predicted that practically no RG ABs should have interacted during their past evolution. Given the short evolutionary timescale of the red-giant-branch (RGB) and later RG stages, their initial masses should therefore be almost equal. So far, only a few ABs have been confirmed or studied as true binary systems \citep{Appourchaux2015,White2017,LiYaguang2018,Beck2018,Marcadon2018, Murphy2021, Grossmann2025, Schimak2026}, but many more AB candidates (ABCs) have been published, investigated, and/or disproven, through kinematics, photometry, and/or seismology \citep[e.g.][and Mazzi et al. in prep.]{Themessl2018b, Bell2019, Espinoza2025, Choi2026}. 

The search effort for evolved ABs is still ongoing, with new techniques under development \citep[see e.g.][]{Choi2025, Choi2026}. Part of the challenge is that current space missions such as \emph{Kepler} and TESS have coarse spatial resolution. This leads to frequent false-positive ABCs due to crowding, where the flux from two neighbouring (on-sky) oscillators of similar apparent magnitude is summed to produce an apparent AB. When these stars can be optically resolved, for example with Gaia DR3 photometry \citep{GaiaDR3}, it is possible to confirm that the two oscillators have different origins on-sky \citep[see e.g.][]{Colman2017, Choi2026}. Techniques involving the detection of a photometric excess can also be used, where astrometric and photometric data of high on-sky resolution is used to measure the absolute flux, which is then compared with an estimate of the intrinsic flux from asteroseismology \citep[][and Mazzi et al., in prep]{Mazzi2025}. For ABs, the detection probability of the dimmer star's oscillations in the power spectrum decreases as the binary luminosity ratio decreases from unity towards zero. Binaries with a detectable second oscillator therefore tend to have higher levels of photometric excess \citep{Mazzi2025}, making the combined photometric and asteroseismic method ideal for searching for new ABCs. However, separating optically unresolved chance alignments from true binaries remains a challenge. In those cases, the detection of Keplerian orbital motion, for example through spectroscopy or astrometry, is the only certain way of confirming that an ABC is in fact a gravitationally bound AB.

Spectroscopic binaries are typically identified through temporal radial velocity (RV) variability (single-lined spectroscopic binary, SB1), and/or through the detection of two RV components (double-lined spectroscopic binary, SB2). The RV semi-amplitude is proportional to $P^{-1/3}$ \citep{Torres2010}, where $P$ is the orbital period. The detected change in RV over a finite baseline therefore decreases with increasing period. For typical oscillating \emph{Kepler} RGs, activity and/or pulsations can complicate the RV measurement up to the 0.1 km/s level \citep{Hekker2008}, but binary RV signals larger than this can still be detected with standard methods. An example of this in the literature is \citet{Price-Whelan2018}.

In this work, our aims are two-fold: to obtain a new sample of confirmed binaries with solar-like oscillating RGs, as well as demonstrate the modelling possibilities offered by ABs. For our first aim, we apply standard RV diagnostics to a sample of ABCs, as well as candidate binaries having only a single detectable oscillator. In Sect.~\ref{sec:observations}, we outline the details of our spectroscopic observing programme of 60 candidates for binary systems, selected based on their seismology (ABCs) and/or based on observed excess photometry compared with expectations from seismology (Mazzi et al., in prep). In Sect.~\ref{sec:binary_sensit}, we quantify the RV sensitivity of our observations as a function of orbital period.  In Sect.~\ref{sec:abs_bf}, we present the analysis of spectral line broadening functions (BFs), which we used to quantify spectroscopic variability in our sample. We summarise the most interesting targets from the spectroscopic study in Sect.~\ref{sec:spec_interesting_targets}. 

Our second aim is to demonstrate the additional modelling possibilities offered by ABs compared to single stars. To this end, in Sect.~\ref{sec:ab_inference_chap4}, we perform a detailed asteroseismic inference for a subsample of binaries and binary candidates, including the peakbagging and modelling of both stars' oscillation mode frequencies in the case of ABCs, in order to demonstrate the benefits of a two-star seismic modelling approach. Finally, we summarise and conclude in Sect.~\ref{sec:conclusion}.
\section{Spectroscopic observations}\label{sec:observations}
\subsection{Campaign summary}
We performed spectroscopic follow-up observations of 60 binary candidates with the high-resolution ($R\sim 67.000$, with full width at half maximum (FWHM) of $ \sim 4.5$ km/s) spectrograph FIES at the Nordic Optical Telescope \citep{Telting2014}, in order to confirm them as spectroscopic binaries. 
With our observational strategy, we aim to obtain at least two spectra with a long baseline, for as many stars as possible, and to follow up with more epochs on the most promising candidate binary systems. For 37/60 of our targets, we obtained at least two spectra, allowing us to investigate potential spectroscopic variability.
This observing proposal was run with filler time status, which meant that observations were only fit into the observing queue on nights with scheduling gaps. For this reason, we kept the exposure times short, aiming for signal-to-noise (S/N) values of 15-25 per spectrum. Due to the low predicted S/N of our observations, we did not attempt a detailed spectroscopic analysis of the individual photospheric lines, instead prioritising the average RV measurement and line shape, where we could effectively stack absorption lines to increase the measurement S/N. Each spectrum had a thorium-argon wavelength calibration spectrum taken immediately before observation. This is typically sufficient for achieving RV precision of 20-40 m/s \citep[see e.g.][]{Thomsen2022, FIES-expmeter, FIES-5ms}. The échelle spectra were reduced by the observatory pipeline, which employs the FIEStool \citep{FIEStool}.

The target selection was an ad hoc mix between several sources of candidates:
\begin{itemize}
    \item 33 targets from an early version of a photometric-asteroseismic catalogue of candidate binaries, based on data from \citet{Willett2026}. This catalogue will be presented in Mazzi et al. (in prep.).
    \item 22 asteroseismic oscillators with poorly fitting single star background fits, which we confirmed through visual inspection as having an additional signal in the power spectrum or peculiar oscillations, making them ABCs (priv. comm., Josefina Montalbán).
    \item 12 ABCs that we found by manually investigating the KEPSEISMIC power spectra \citep[e.g.][and references therein]{Pires2015}.
    \item 13 targets from the samples of ABCs and peculiar oscillators presented in \citet{Colman2017,Bell2019,Gaulme2020,Schonhut-Stasik2020}.
\end{itemize}
There is overlap between these sources. The Tables \ref{table:spec_abs} and \ref{table:spec_1spec} in the appendix highlight the source(s) for each target, as well as the number of observations for each. For the 37 of 60 stars for which we obtained more than one spectrum, the typical number of observations was 2. Our targets were selected from a larger preliminary collection, after cross-matching with the Gaia DR3 catalogue to remove targets with a potential contaminating star, since those could be due to a chance alignment \citep[see e.g. ][]{Miglio2014, Colman2017}. Due to this, our target sample is unlikely to be complete. However, this significantly helped reduce observing time requirements. Of the 60 targets in our sample, only KIC\,4346953, 5015647, 6501237, 12117790, 9392647, and 9392650 had a nearby ($< 10^"$) contaminant less than 3 magnitudes dimmer. These targets' contaminants are investigated in detail in Appendix~\ref{sec:spec_interesting_targets}, when relevant.
\subsection{Sensitivity to binarity}\label{sec:binary_sensit}
True binarity requires that the stars are gravitationally bound and orbit each other. From the population synthesis simulation of \citet{Mazzi2025} using the \citet{MoeStefano2017} binary parameter distributions, $\sim20$\% of RG ABs are expected to have orbital periods below $10^3$ days, $\sim40$\% below $10^4$ days, and $\sim 60\%$ below $10^5$ days. For the 37 stars where we have at least two spectra, our observational baseline spans between 42 and 413 days. We here present a simplified calculation of the expected RV signals of gravitationally bound binaries, to simulate the detection sensitivity of our observational strategy as a function of orbital period. We remark that the goal of our campaign was to confirm as many true binaries as possible, for the purpose of future study of those systems, and not to obtain a complete account of all ABs in our sample, nor in the full \emph{Kepler} RG sample.

For each orbital period evaluated, we simulated 10,000 gravitationally bound orbits, assuming that both stellar components have $M=1\textrm{M}_{\odot}$ ($M_2/M_1 = 1$). We assumed an isotropic distribution of orbital inclinations (a uniform distribution in $\cos i$ between $[0, 1[$), and a uniform distribution in periastron longitude between $[0^{\circ}, 360^{\circ}[$. For the eccentricity, $e$, we assumed the solar-type distribution from \citet{MoeStefano2017}, with $p_e \propto e^{0.4}$ and $e < 1-(P/(2 \textrm{ days}))^{-2/3}$, where $P$ is the orbital period in days. We verified that the results do not change noticeably if we adopted a uniform eccentricity distribution between $]0,1[$. Finally, we assumed a uniform distribution for the time of first observation, between $[t_{0,\rm peri}, t_{0,\rm peri}+P[$, where $t_{0,\rm peri}$ is the time of periastron passage. For each simulated orbit, we then evaluated the component RVs assuming that the target was observed only two times, with a separation between consecutive observations (baseline) of either 50 or 400 days. These baselines are approximately representative of the upper and lower limits for our campaign.

For targets where one star produces most of the flux, the variation in RV of the brightest component can approximately describe the binary detection probability. For the other extreme, where the binary luminosity ratio is close to unity, the spectroscopic profile will be clearly resolvable as bimodal only if the difference in RV ($rv_1-rv_2$) is larger than the FWHM of the two stellar components' absorption profiles. With low RV difference, the composite profile will be indistinguishable from a typical single-lined absorption profile, only with a slightly larger width. In that case, a change in $rv_1-rv_2$ will modify the apparent width of this composite profile, which we approximate using a Gaussian function with modified $\sigma = \sqrt{\sigma_{\textrm{init}}^2+(rv_1-rv_2)^2}$, where $\sigma_{\textrm{init}}$ is the original width of a single component.

\begin{figure}
    \centering
    \includegraphics[width=\columnwidth]{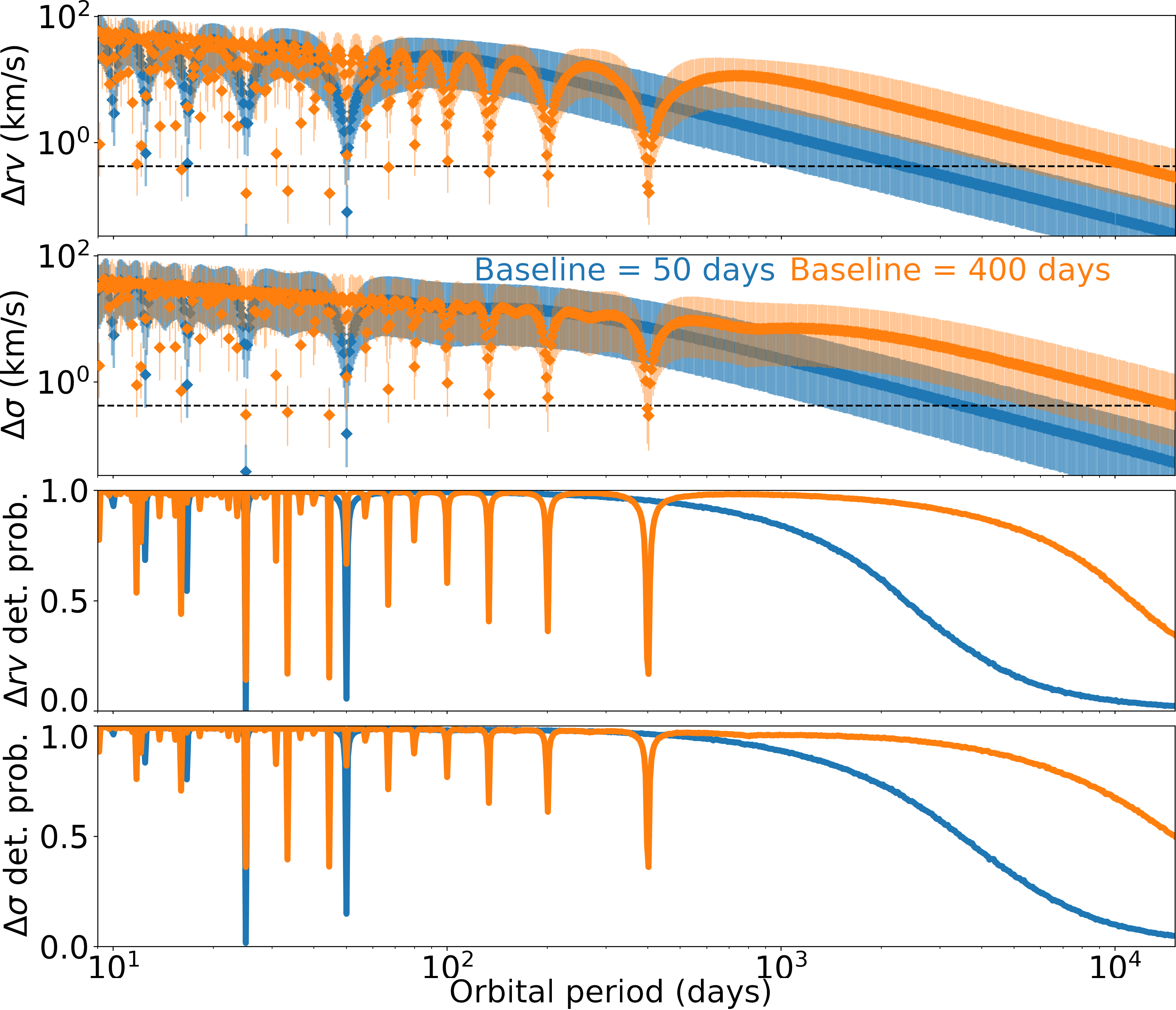}
    \caption{Simulated observations of spectroscopic variation due to Keplerian motion for bound binaries of varying orbital periods, assuming only two observations separated by 50 (blue) or 400 (orange) days. See the text for a description of $\Delta rv$ and $\Delta \sigma$. Upper panels show 68\% confidence intervals for the predicted variability (the error bars). The medians are marked by diamonds. The black dashed line is the assumed 0.42 km/s limit for a 99.7\% confidence detection (see text). Bottom panels show 99.7\% confidence detection probabilities assuming the 0.42 km/s threshold.}
    \label{fig:rv_simulation}
\end{figure}

In Fig.~\ref{fig:rv_simulation}, we present these variation measures for the simulated binaries; either the RV variation $\Delta rv$ of a spectroscopic profile dominated by a single star, or the change in apparent width $\Delta \sigma$ of an unresolved combination of two stellar components' absorption lines with equal luminosity ratio. For calculating $\Delta\sigma$, we assumed that the width of the spectroscopic profiles is dominated by the instrumental broadening of FIES ($\textrm{FWHM} \sim 4.5$ km/s) rather than stellar surface rotation. For most RGs, this is a reasonable assumption \citep[e.g. ][]{Ceillier2017}.

To estimate a detection probability, we assumed a limiting RV precision of 0.1 km/s per spectrum, which is a reasonable limit when considering the typical astrophysical variability of RGs with $\log g > 2$ \citep{Hekker2008}. If we obtained just two spectra, the $99.7\%$ confidence detection limit for $\Delta rv$ is therefore $0.42$ km/s. For the $\Delta \sigma$ estimate, the role of astrophysical variability is less obvious to estimate. We therefore conservatively assumed the same limit for simulating the detection probability. We find that both measures remain sensitive to the signal induced by a binary orbit, even for periods above $10^4$ days for a 400 day baseline. For binaries with period 10-100 days, we predict an integrated average detection probability of $95\%$ ($98\%$) for the 400 (50) day baseline. For 100-1000 days this is $96\%$ ($93\%$), and for 1000-10,000 days we predict an average detection probability of $80\%$ ($24\%$). If all observations were taken with a baseline of 400 days, we estimate an approximate completion factor of $\sim 40\%$ for the confirmation of ABs of unknown orbital periods, based on our simulated detection probabilities and the period distribution of simulated ABs in \citet{Mazzi2025}.
\section{Broadening function analysis}\label{sec:abs_bf}
We aim to derive spectroscopic measures of binarity. In order to conclusively confirm binarity spectroscopically in this work, we require that they have significant temporal variability of the average spectroscopic line profile (see Sect.~\ref{sec:binary_sensit}). The measures required to confirm binarity are outlined in Sect.~\ref{sec:binary_multi}.
For all stars, we furthermore estimated the width and skew of the average line profile (see Sect.~\ref{sec:binary_single}), as these can be an indicator of binarity. In Appendix~\ref{sec:spec_interesting_targets}, we summarise key details about specific targets of interest in our observational campaign.

Before detailing the binary indicators based on the line profile, we first outline the calculation of the average line profile, and considerations for RV fitting and uncertainty estimation.
We performed dimensionality reduction on the wavelength spectra by calculating the average line BF \citep{Rucinski1999} of each spectrum. The BF is obtained by least-squares deconvolution of the observed spectrum using a sharp-line template spectrum. It is therefore a best-fit representation of the average shape and Doppler-shift of the spectral lines. Large asymmetries, bi-modality, as well as changes in BF shape and RV over time can be indicators of binarity, which can be explored with model profile fitting. A double-lined binary (SB2) will have a bimodal line profile that varies following a Keplerian orbit of the two stellar components, while a single-lined binary (SB1) will have a unimodal line profile following only the orbit of the brighter star. A static but bimodal BF can indicate a chance alignment of two stars that are not optically resolved, but this can require a long time baseline to verify unless the stars are well separated in RV. The detection of binarity through spectroscopic variability measurements such as ours was illustrated in Sect.~\ref{sec:binary_sensit}.

We used a fixed sharp-line synthetic template spectrum for all stars from \citet{Coelho2005}, with $T_{\rm eff}=4750$K, $\log g=2.5$, $\rm [Fe/H] = 0.0$, and $[\alpha/\textrm{Fe}] = 0.0$, and limited the calculation to the optical region $4500-5825$\AA~to avoid telluric lines and interstellar Na in the observations, as well as a low S/N in the blue. We further smoothed the BFs with a Gaussian kernel having $\sigma=1$km/s to reduce regression-induced noise from the wavelength-to-velocity transformation.

To quantify spectroscopic binarity of the targets, we performed model fits to the BF of each spectrum. We used a Gaussian model unless otherwise stated. With this fit we assumed that the target is single-lined and not significantly broadened by rotation. The latter is a reasonable assumption for most RGs, because the line profile is often dominated by instrumental broadening due to slow surface rotation \citep[see e.g.][]{Ceillier2017}. The Gaussian model is
\begin{align}
    f_{\rm Gauss}(v) &= \frac{a}{\sigma\sqrt{2\pi}}e^{\frac{-(v-rv)^2}{2\sigma^2}},
\end{align}
with $a$ and $\sigma$ being the amplitude and standard deviation of the Gaussian function, respectively,  $v$ the velocity bin, and $rv$ the RV centre of the Gaussian.

For parameter fitting and uncertainty estimation, we used non-linear least squares optimisation. The fitted-parameter uncertainties are formal least-squares estimates, whose reliability depends on the uncertainties assigned to the input BF values. We estimated the uncertainty of the BF data from the variance outside of the line profile. Later in this section, we evaluate whether this estimate is sufficient.

In cases where the line profile was clearly bimodal, we fitted a combined two-Gaussian model. Here, we fitted all the BFs simultaneously, to enforce the same line widths and amplitude ratios. We performed a linear least squares fit within the non-linear optimisation, in order to marginalise out the continuum and absolute amplitude of each BF. This way, the non-linear optimisation had $3+2n$ free parameters ($a_2/a_1$, $\sigma_1$, $\sigma_2$, $n \times rv_{1}$, and $n \times rv_{2}$), where $n$ is the number of spectra. In Table~\ref{table:spec_abs}, we call this bimodal model ‘SB2’, and the corresponding unimodal model ‘SB1’. The use of this model does not inherently mean that the stars are true binaries, just that we fitted them with a double- or single-lined profile.

The data in the BFs include a significant level of correlation (sinusoid-like features in Fig.~\ref{fig:2971380_bf}). Some of this correlation is systematic and due to a mismatch between some of the features of the template and the observed spectrum. However, the majority comes from the transformation of a noisy spectrum from wavelength to velocity space \citep[see ][]{Rucinski1999}. This can also be considered over-fitting noise, since it is amplified by the size of the BF design matrix. We have attempted to dampen this noise through smoothing with a Gaussian filter having $\sigma = 1 $km/s. This reduces the amplitude, but also imprints an additional correlation between the velocity bins. For KIC\,2971380, which is the only binary where we had sufficient spectra to model the full orbital motion, we explored the impact of the total correlation on the output uncertainties. We assumed that the residuals follow a multivariate normal distribution with a stationary covariance; that is, one that is dependent on only the velocity difference between two bins, and not the absolute velocity. We estimated this covariance matrix by calculating the autocorrelation function of the residuals, after a best-fitting BF model was subtracted. We found that the resulting orbital parameters and uncertainties of KIC\,2971380 are not significantly different from those obtained when neglecting this correlation. The output parameters for KIC\,2971380 presented in this work were calculated including this contribution. However, based on these findings we have neglected the correlation for the targets in the sample that have fewer spectra, in order to preserve computational efficiency.
\subsection{Binary signature - Multiple nights}\label{sec:binary_multi}
Below, we detail the variability flags used to quantify binarity. We observed 37 stars with observations over at least two nights. For these stars, we explored if temporal variations were detectable in the BF. These are summarised in Table~\ref{table:spec_abs} in the appendix. The measures quoted in the table are outlined below.
\begin{itemize}
    \item Slope, $\frac{\textrm{d}v}{\textrm{d}t}$: The RV slope of a linear fit of the RVs as a function of time, thereby assuming that the candidate binary has a period significantly longer than the baseline.
    \item Slope fit, $\Delta$BIC: The fit quality of the above line fit against a constant fit, by comparing the Bayesian information criterion (BIC) of both. Negative $\Delta$BIC means that the line fit was preferred over the constant.
    \item RV scatter, $\Delta v$: The RV variation, estimated as the combination of the two RV measurements with the highest significant difference relative to their quadrature-combined uncertainty.
    \item Change in width, $\Delta \sigma$: The most significant variation in Gaussian width, $\sigma$, by fitting a single Gaussian profile to the BF.
\end{itemize}
We added 100 m/s in quadrature to all RV and $\sigma$ uncertainties, estimated to account for potential astrophysical variability in the line profile (see Sect.~\ref{sec:binary_sensit}).

For ten of the 37 systems, the BF profile was visually bimodal, indicating the presence of two resolved stellar absorption components.
We therefore fitted a combination of two Gaussians, with the same method as described for KIC\,2971380 in Sect.~\ref{sec:abs_bf}. We verified that these two-profile models had a better BIC than a one-profile fit.
For these ten systems, the velocity values reported in Table~\ref{table:spec_abs} refer to the separation $RV_2-RV_1$, rather than the absolute velocities.
For example, the parameter $v$ is the most significant difference of all the measurements of $|RV_2-RV_1|$.
These double-lined measurements are independent of any potential drift in the instrumental zero-point, but they can still contain astrophysical variability of the line profile.

In order to report a star as significantly variable in Table~\ref{table:spec_abs}, we required that $\Delta\textrm{BIC} < -10$ (SB1), $\Delta\textrm{BIC} < -2$ (SB2), time variation in the width ($\Delta\sigma > 3\cdot \textrm{Err}(\Delta\sigma)$), or time variation in the RV ($\Delta v > 3\cdot \textrm{Err}(\Delta v)$).
A few stars show tentative indications of binarity, either $\Delta\sigma/ \textrm{Err}(\Delta\sigma) \sim 2-3$, or both the slope and $\Delta v$ approximately similar to the uncertainty.
In particular, KIC\,11299484 shows a clearly bimodal line profile, but the RV variation is only at the $1\sigma$ significance level over a baseline of 255 days.
In total, we detect highly significant variability ($>99.7\%$ confidence) for 12 systems, 8 of which we further confirm as SB2. 8/12 systems are ABCs. We detect variability at marginal significance for another 7 systems (all ABCs).
\subsection{Single spectrum indicators}\label{sec:binary_single}
\begin{figure}
    \centering
    \includegraphics[width=\columnwidth]{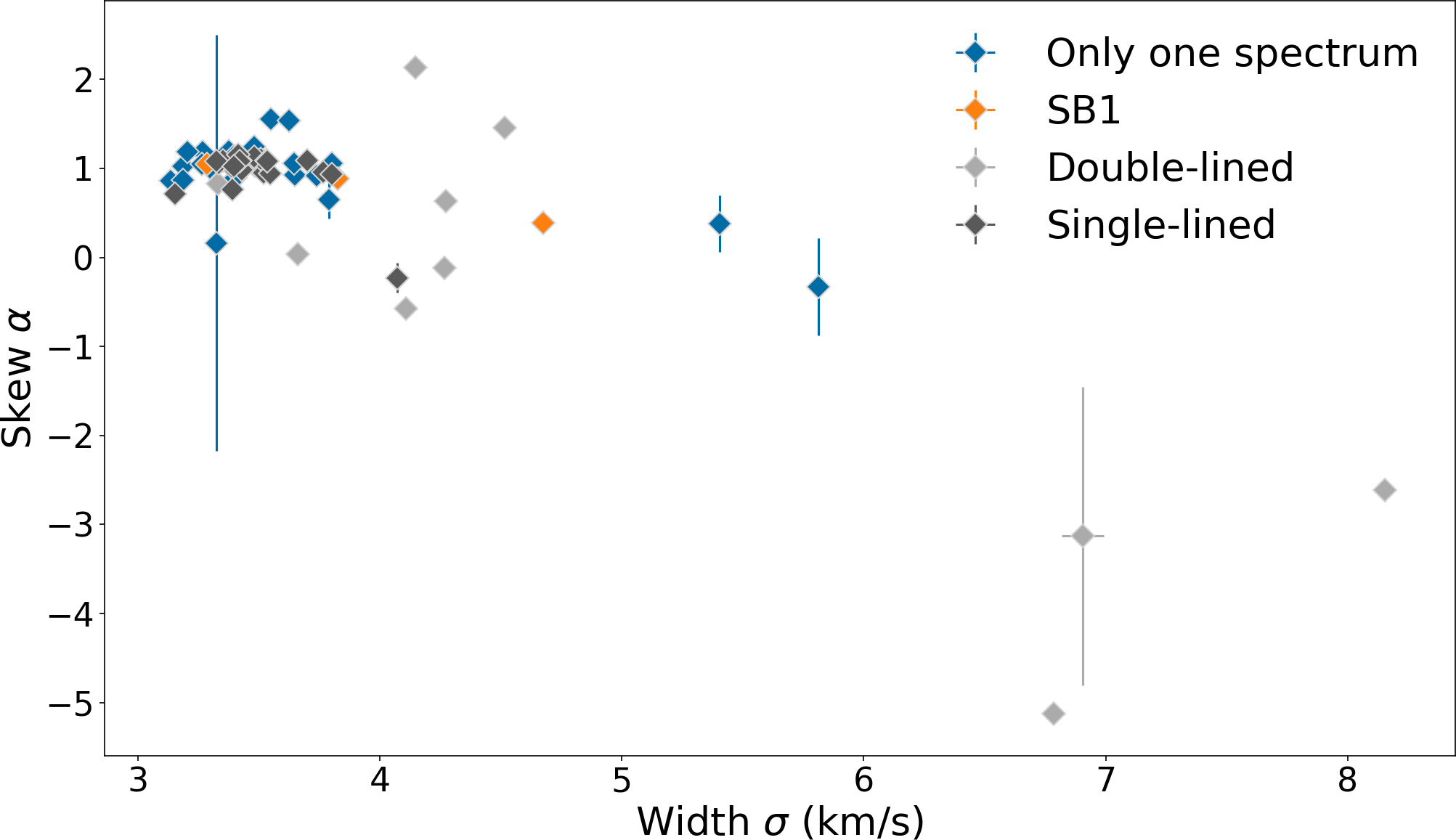}    \caption{BF asymmetry, $\alpha$, versus BF width, $\sigma$. Blue: Stars with only one night's observation. Orange: RV variable single-lined binaries. Light grey: Double-lined stars (both variable and non-variable). Dark grey: Single-lined stars with multiple nights of observations.}
    \label{fig:sigma_skew}
\end{figure}

For 23 out of 60 targets, we only have FIES observations from a single night. For those, it is not possible to conclusively confirm binarity through RV variability. SB2 systems, however, might still show significant deviations in the line profile relative to what is expected for a RG. We quantify this through two measures: the width, $\sigma$, of a Gaussian fitted to the observed BF, and the asymmetry factor, $\alpha$, of an asymmetric Gaussian model \citep{Azzalini1985}: 
\begin{align}
    f_{\rm asym}(v) &= f_{\rm Gauss}(v) \cdot \left[1+\mathrm{erf}\left(\frac{\alpha(v-rv)}{\sigma\sqrt{2}}\right)\right].
\end{align}
\noindent The asymmetry factor is not expected to be centred on zero for real stars, since the FIES instrumental profile is asymmetric (priv. comm., John Telting) and the stellar lines are affected by convective blueshift \citep{Gray2009}. The two measures are shown in Fig.~\ref{fig:sigma_skew} and Table~\ref{table:spec_1spec} for the whole sample. This way, we can compare the single-epoch stars to the SB1 variable stars, the stars with a clear bimodal profile and multiple spectra (‘double-lined’), and the single-lined stars that have multiple observations. The latter category can in principle serve as a control sample for the expected profile shape for single stars. See Sect.~\ref{sec:binary_multi} for the determination of RV variability.

From our measures, it appears that 5-6 single-observation stars are outliers compared to the typically expected skewness ($\sim 1$) and width ($3-3.8$ km/s). These are KIC\,3860550, 3955502, 4058578, 5700488, 6706199, and potentially 4282111.
Qualitatively, we further remark that the BF profile of KIC\,3860550 has clear visual indications of being bimodal.

\subsection{Binary indicators - APOGEE and Gaia}\label{sec:binary_gaia}
We cross-matched our sample with Gaia DR3 \citep{GaiaDR3} and APOGEE DR17 \citep{apogeedr17}, to compare with variability measures from the two catalogues. In Table~\ref{table:spec_1spec}, we have summarised the Gaia DR3 astrometric excess measure RUWE (renormalised unit weight error), the Gaia RV variability flag (called Var/DR3) following \citet{Katz2023}, obtained by requiring rv\_chisq\_pvalue $< 0.01$ and rv\_renormalised\_gof $> 4$. We also include the non-single-star (NSS) flag \citep{GaiaDR3NSS}, the maximum Gaia RV amplitude after outlier removal \citep{Katz2023}, and the APOGEE DR17 RV scatter \citep{apogeedr17}. There is overlap between ours and the catalogue determinations, particularly for the single-lined RV variable stars. The survey indicators are less clear for the SB2 systems, likely due to the lower spectral resolution employed.

Based on this cross-comparison (see Tables~\ref{table:spec_abs},\ref{table:spec_1spec}) for the targets that are unconfirmed and have one or more spectra, we added the following sources as likely binaries: KIC\,4282111, 4637699 (ABC), 8181509, 8295007 (ABC), 10841730 (ABC), 11299484 (ABC), 11352644 (ABC), 11465942, and 12117790 (ABC). These have either one clear signature in Gaia/APOGEE, or tentative indications of binarity from two or more sources. Note in particular that KIC\,10841730 is a confirmed AB \citep{Schimak2026}, and KIC\,6206407 was already identified in \citet{Espinoza2025} as an ABC with a high RUWE value.
\section{Asteroseismic inference of binaries}\label{sec:ab_inference_chap4}
In this section, we present detailed peakbagging and asteroseismic inference of one ex situ (accreted) Galactic binary for which we measured the binary mass ratio with RVs in Sect.~\ref{sec:abs_bf} (KIC\,2971380), and a proof-of-concept seismic investigation for one confirmed AB (KIC\,10592924) and two ABCs (KIC\,6501237 and 11299484), with the aim of using the oscillations of both stars. We selected the AB and ABCs based on a preliminary visual inspection of their power spectra, where we confirmed based on their mixed modes \citep{Bedding2011} that they were likely to contain two RGB stars with oscillations well separated in frequency. The two ABCs were also selected to investigate with asteroseismology if their two oscillators were compatible with having formed as a binary or part of a common stellar population.
\subsection{Oscillation frequency measurement}\label{sec:specseis_peakbag}
To measure the observed oscillation frequencies in the KEPSEISMIC power spectra \citep[e.g.][and references therein]{Pires2015}, we performed peakbagging by fitting a Lorentzian profile to each observed radial ($l=0$), dipole ($l=1$), and quadrupole ($l=2$) mode. The aim was to accurately measure the radial mode frequencies for asteroseismic inference, and in cases where dipole modes were challenging to model with a single Lorentzian profile, we excluded them from the fit.

First, we fitted an asteroseismic background and power excess with DIAMONDS \citep{Corsaro2014} to capture granulation, oscillation power excess, and white noise, $W$. For the granulation signal of KIC\,2971380, we used three super-Lorentzian profiles \citep{Kallinger2014}. For ABs, we implemented a model with two super-Lorentzian profiles and two Gaussian power excesses:
\begin{align}
        \mathcal{P}(\nu) &= \eta(\nu)\left[\sum_{k=1}^2 \left(\frac{2\sqrt{2}\sigma_{k}^2}{\pi\nu_{k}\left(1+\left(\frac{\nu}{\nu_k} \right)^4\right)}\right)
        + \sum_{j=1}^2 \left(H_{\textrm{ osc},j} e^{-\frac{(\nu-\nu_{\textrm{ max},j})^2}{2\sigma_{\textrm{ osc},j}^2}} \right)\right] + W, \nonumber\\
    \eta(\nu) &= \mathrm{sinc}\left(\frac{\nu}{2\nu_{\rm nyquist}}\right)^2.\label{eq:background}
\end{align}
$\eta$ is the attenuation function accounting for the finite integration time of \emph{Kepler}.
We attempted to use more granulation terms to match both stars in the ABs, but these could not be constrained independently. When we attempted to fit the low-frequency power with a third granulation component, we had issues with getting DIAMONDS to complete the sampling. For the ABs, we excluded the data from 0-4 $\mu$Hz, while for KIC\,2971380 we only excluded data below the window frequency of the KEPSEISMIC light curve filter (20 days, $\sim 0.6 \mu$Hz).

After background fitting, we performed peakbagging of KIC\,2971380 with DIAMONDS. The $\ell = 1$ modes had a well-resolved mixed mode characteristic, and we therefore fitted the individual mixed modes with narrow Lorentzians as well, since they were not fully unresolved as sinc profiles.

For the ABs, we peakbagged the oscillation modes with a custom-made script using the affine-invariant Markov chain Monte Carlo (MCMC) sampler \textsc{emcee} \citep{foreman-mackey2013}. This was necessary as current automatic peakbagging codes \citep[e.g. PBJam, DIAMONDS, FAMED][]{Nielsen2021, Corsaro2014, FAMED} are not designed to handle two oscillation envelopes. For the peakbagging we also accounted for the impact of the integration time on the oscillation amplitude. When evaluating the likelihood, we assumed that the noise characteristics of the power spectrum follow a $\chi^2$ distribution with two degrees of freedom.

The individual oscillation frequencies are reported in Table~\ref{table:ab_freqs}.
The observed $\ell = 1$ modes of KIC\,2971380 are reported in Table~\ref{table:29_l1}.
In addition to this, we measured the large frequency separation between radial orders, $\Delta\nu_0$ (Table~\ref{table:ab_obs_seis}), by performing a linear fit of the radial mode frequencies, weighting the modes by the inverse of the observed variance.
The reported uncertainties for $\Delta\nu_0$ are only the formal uncertainties.
We also used the observed asymptotic relation of \citet{Mosser2011,Mosser2013} to measure $\Delta\nu_{\rm obs}$, small frequency separation, $\delta\nu_{\rm obs,02}$, offset, $\epsilon_{\rm obs}$, and curvature, $\alpha_{\rm obs}$.
Specifically, we used the relation
\begin{align}
    \nu_{n,l} &= \left(n+\epsilon_{\rm obs}+ (\alpha_{\rm obs}/2)\left( n-\frac{\nu_{\rm max}}{\Delta\nu_{\rm obs}} \right)^2 \right) \Delta\nu_{\rm obs} ~~+\delta_{\textrm{obs}, 0l}.
\end{align}
Here, we sampled the probability distribution with MCMC, including the observed $\nu_{\rm max}$ (with uncertainty).
We furthermore included a nuisance parameter, $\sigma_{\rm jitter}$, to account for intrinsic scatter, inflating the uncertainties by $\sigma = \sqrt{\sigma_{\rm obs}^2+\sigma^2_{\rm jitter}}$, since the detailed structure of the mode frequencies is not modelled with the asymptotic relation. This resulted in significantly higher parameter uncertainty estimates. The quoted uncertainties for $\Delta\nu_{\rm obs}$ in Table~\ref{table:ab_obs_seis} are therefore more realistic than those reported for $\Delta\nu_0$.
\subsection{Asteroseismic modelling inference}\label{sec:abs_inference}
We performed an asteroseismic inference using a grid of stellar models with calculated theoretical oscillation frequencies.
We used the alpha-version of AIMS3\footnote{\url{https://gitlab.com/warrickball/aims3-alpha}}, which is a partial re-implementation of the Bayesian inference code AIMS \citep[Asteroseismic Inference on a Massive Scale, ][]{Rendle2019}. AIMS3 allowed us to define a likelihood function including two stars, whereas we could only study one star at a time in the previous version. For mode identification, we used the older version of AIMS, since this was not reimplemented in AIMS3.

We used a grid of stellar evolutionary tracks and theoretical frequencies calculated with MESA version 11701 \citep[][and references therein]{paxton2019} and GYRE \citep{townsend2013}. This grid will be presented in full detail in Tailo et al. (in prep.), but it has already been tested on the dEB KIC\,10001167 \citep{Thomsen2025}. Each model was evolved from the pre-main-sequence until the first thermal pulse. However, we restricted the grid to the sub-giant and RGB phases when fitting. The full grid spans: [Fe/H] from -2.0 to +0.55 with typical step of 0.25 dex, with solar mixture from \citet{Asplund2009}; [$\alpha$/Fe] from -0.2 to 0.4 in steps of 0.2 dex; initial helium abundance, $Y_i$, from 0.22 to 0.34 in steps of 0.03; and masses from 0.6 to 1.8 in steps of 0.05 $\textrm{M}_{\odot}$, with additional models at mass $2.0, 2.15, 2.3, 2.45, 2.75, 3.0 \textrm{M}_{\odot}$. The atmosphere description by \citet{KrishnaSwamy1966} was used, and we included microscopic diffusion. Further choices regarding the equation of state, opacities, and mixing were the same as in \citet{Miglio2021}. The mixing-length theory (MLT) was chosen to describe convection with the formula from \citet{cox1968}, and $\alpha_{\rm MLT}=2.2902$ was obtained by solar calibration.

As observational constraints in AIMS3, we used the radial mode oscillation frequencies and the frequency of maximum oscillation power, and the observed surface metal fraction, $(Z/X)_{\rm surf}$, obtained from the iron abundance, [Fe/H], combined with [$\alpha$/Fe], both from APOGEE DR17 \citep{apogeedr17}. We converted to $Z/X$ with
\begin{align}
    (Z/X)_{\textrm{surf}} &= (Z_{\odot}/X_{\odot}) \cdot 10^{[\textrm{M}/\textrm{H}]}, \\
    [\textrm{M}/\textrm{H}] &= [\textrm{Fe}/\textrm{H}] + \log_{10}(0.6387\cdot 10^{[\alpha/\textrm{Fe}]} + 0.3613). \nonumber
\end{align}
The [M/H] calculation is based on the \citet{Salaris1993} correction, using the \citet{Asplund2009} solar mixture.
Since all the stars we analysed are in binary systems, we only applied the metallicity constraint on the brightest star. The metallicity could be biased by dilution from the companion and blending of the lines. 
In Appendix~\ref{sec:spec_binary_bias}, we confirm that the expected bias on the spectroscopy from the presence of two overlapping sources is at maximum 0.1 dex for our targets, and we therefore adopted this limit for the metallicity uncertainty.

We corrected theoretical frequencies for the surface effect (stemming from model inaccuracies near the surface) using the cubic (one-term) prescription of \citet{ballgizon2014} unless otherwise stated, with a correction for each star. These are included as additional sampling parameters. For KIC\,6501237, the fit quality was much better when using the two-term prescription of \citet{ballgizon2014}, including an inverse term \citep[see also][]{Joergensen2021}. The two-term prescription is
\begin{align}
    \delta\nu_{\textrm{surf}}(\nu, \mathcal{I}) = \left[a_{-1}(\nu/\nu_{\rm ac})^{-1}+a_3(\nu/\nu_{\rm ac})^3\right]/\mathcal{I},
\end{align}
\noindent  while the cubic (one-term) prescription excludes the $a_{-1}$ term. Here, $\nu$ is a theoretical mode frequency, $\nu_{\rm ac}$ is the acoustic cut-off frequency, and $\mathcal{I}$ is the normalised mode inertia. As a proxy of $\nu_{\rm ac}$, we used the model $\nu_{\textrm{max}}$. Since $\nu_{\rm ac}$ and $\nu_{\textrm{max}}$ are approximately proportional \citep{Brown1991, Belkacem2011}, this replacement simply results in a rescaling of $a_3$ and $a_{-1}$.

We defined a logarithmic likelihood function comprised of the following $\chi^2$ terms for the brighter (bri) and dimmer (dim) stellar components in the ABs:
\begin{align}
    -\ln \mathcal{L} &\propto \chi^2_{\textrm{bri}, \ell=0} + \chi^2_{\textrm{dim}, \ell=0} + \chi^2_{\textrm{bri}, \nu_{\rm max}} + \chi^2_{\textrm{dim}, \nu_{\rm max}} + \chi^2_{\textrm{bri}, (Z/X)},\\
    \chi^2_{\ell=0} &= \sum_n \frac{(\nu_{n,0} - (\nu_{n,0\textrm{ model}}+\delta\nu_{\textrm{surf }n,0}))^2}{\sigma_{n,0}^2},\\
    \chi^2_{\nu_{\textrm{max}}} &= \frac{(\nu_{\textrm{max}} - \nu_{\textrm{max, model}})^2}{\sigma_{\nu_{\textrm{max}}}^2},\\
    \chi^2_{(Z/X)} &= \frac{((Z/X)_{\textrm{surf}}-(Z/X)_{\textrm{surf, model}})^2}{\sigma_{(Z/X)_{\textrm{surf}}}^2}.
\end{align}
With this likelihood, we do not include any relative weighting of the ‘classical’ and ‘seismic’ constraints, since our only non-seismic constraint is the metallicity, which is only conservatively constrained with an uncertainty of 0.1dex on [Fe/H].

For sampling the likelihood, we used MCMC with \textsc{emcee} \citep{foreman-mackey2013}.
We used an RGB sub-grid of the MESA grid, with fixed $[\alpha/\textrm{Fe}]=0.0$ for the ABs, and $[\alpha/\textrm{Fe}]=0.15$ for KIC\,2971380. We sampled the following stellar parameters: the mass of the brightest star, mass ratio of the binary, shared metallicity, [Fe/H], shared initial helium fraction, $Y_i$, and a-dimensional age, $\tau$ (one for each star). 
We define $\tau$ as the mass of the helium core, which is a feature that increases monotonically with real age, $A$, during the full RGB. We used $\tau$ in place of the real age, both for sampling and interpolation, in order to reduce the occurrence rate of interpolation failure during sampling \citep[see][]{Rendle2019}. We verified that this parameter transformation did not bias the likelihood sampling, since the logarithm of the Jacobian (the logarithm of the partial derivative $\ln (\partial \tau/\partial A)$) is well approximated as locally constant relative to the overall gradient of the likelihood.

We sampled an a-dimensional age, $\tau$, for each star, but enforced coevality by including a Gaussian prior on the age difference of the stars, requiring that their real ages agree within $5\%$ of the mean age of the two stars (e.g. 250 Myr for 5 Gyr). This is roughly $50-90\%$ of the typical statistical uncertainty on age that we expect for a single RGB star with \emph{Kepler} data on individual mode frequencies \citep[see e.g.][]{Montalban2021, Thomsen2025}, and it is therefore not a particularly restrictive constraint. The agreement (or disagreement) between the posterior ages of the two stars, together with the seismic fit quality, can therefore further inform us of whether coevality is a good assumption.
\begin{table*}
    \centering
\caption[Asteroseismic model inferences for four binaries.]{\footnotesize Asteroseismic radial mode model inferences\tnote{a}.}\label{table:ab_inference}
\begin{threeparttable}
    \footnotesize
\begin{tabular}{lccccc}
    KIC & & 6501237 & 10592924 & 11299484 & 2971380\\
    \hline
    Input: $(Z/X)_{\rm surf, bright}$\tnote{b} &    & 0.0152 (35)   &0.0096 (22) & 0.0191 (44)   & 0.0039 (13)\\
    Input: $\nu_{\rm max, bright}$ &$\mu$Hz         & 62.86 (19)    &52.52 (24) &48.27 (0.23)     & \\
    Input: $\nu_{\rm max, dim}$    &$\mu$Hz         &123.18 (41)    &130.56 (82) & 89.82 (0.41)    &\\
    Input: $[\alpha/\textrm{Fe}]$    &dex         &0.0    &0.0    & 0.0 &0.15\\
    Input: $M_{\textrm{dim}}/M_{\textrm{bright}}$    &         &    &    &&1.005 (10)\\
    Input: $L_{\textrm{dim}}/L_{\textrm{bright}}$    &         &    &    &&0.25 (5)\\
    \hline
    $M_{\rm bright}$ &$\textrm{M}_{\odot}$           & 1.167 (21)    &1.223 (18)     &1.200 (18)  & 0.871 (17)     \vspace{0.015in}\\
    $M_{\rm dim}$ &$\textrm{M}_{\odot}$              &1.165 (20)     &1.215 (19)     &1.203 (19)  &0.865 (17)\vspace{0.015in}\\
    $M_{\textrm{dim}}/M_{\textrm{bright}}$& & 0.9980 (91)   &0.993 (10)     & 1.001 (10) &0.9940 (29)               \vspace{0.015in}\\
    $\textrm{[Fe/H]}$ & dex                 & -0.10 (10)    &-0.322 (78)    & -0.038 (83)& -0.752 (86)   \vspace{0.015in}\\
    $Y_i$              &                    &0.296 (11)    &0.2667 (84)     & 0.2720 (86)&0.25154 (89)\tnote{c}\vspace{0.015in}\\
    $\textrm{Age}_{\rm bright}$ &Gyr        & 4.65 (49)     &4.00 (32)      & 5.47 (61)  & 11.70 (99)     \vspace{0.015in}\\
    $\textrm{Age}_{\rm dim}$ &Gyr           &4.63 (52)      &4.03 (34)      & 5.41 (62)  &\vspace{0.015in}\\
    $R_{\rm bright}$ &$\textrm{R}_{\odot}$           & 7.975 (56)    &8.930 (42)     & 9.228 (48)& 6.167 (40)    \vspace{0.015in}\\
    $R_{\rm dim}$ &$\textrm{R}_{\odot}$              &5.645 (35)     &5.562 (31)     & 6.751 (38)& 2.80 (32)\vspace{0.015in}\\
    $\log g_{\rm bright}$ & dex             & 2.7020 (21)   &2.6238 (21)    & 2.5871 (22)& 2.7981 (27)   \vspace{0.015in}\\
    $\log g_{\rm dim}$ &  dex               &3.0012 (22)    &3.0320 (21)    & 2.8595 (21)& 3.48 (10)\vspace{0.015in}\\
    $\rho_{\rm bright}$& $10^{-3}\rho_{\odot}$& 2.301 (10)  &1.7173 (14)    & 1.5265 (25)& 3.7151 (70) \vspace{0.015in}\\
    $\rho_{\rm dim}$ & $10^{-3}\rho_{\odot}$& 6.473 (25)    &7.0596 (78)    & 3.9118 (44)& 39 (15)               \vspace{0.015in}\\
    $L_{\rm bright}$ &$L_{\odot}$           & 31.6 (14)     &40.7 (14)      & 38.5 (16)  & 21.06 (72)     \vspace{0.015in}\\
    $L_{\rm dim}$ &$L_{\odot}$              &17.46 (78)     &17.99 (62)     & 22.63 (95) & 5.1 (10)\vspace{0.015in}\\
    $(Z/X)_{\rm surf, bright}$ &            & 0.0152 (31)   &0.0091 (15)    & 0.0173 (32)& 0.00411 (90)   \vspace{0.015in}\\
    $(Z/X)_{\rm surf, dim}$ &               &0.0151 (31)    &0.0091 (15)    & 0.0173 (32)& 0.00390 (70)\vspace{0.015in}\\
    $T_{\rm eff, bright}$ &K                & 4848 (42)     &4880 (32)      & 4733 (38)  & 4982 (35)     \vspace{0.015in}\\
    $T_{\rm eff, dim}$ &K                   &4967 (43)      &5041 (33)      & 4846 (40)  & 5178 (58)\vspace{0.015in}\\
    $\nu_{\rm max, bright}$ &$\mu$Hz        & 62.77 (15)    &52.29 (16)     & 48.38 (14) & 76.79 (41)    \vspace{0.015in}\\
    $\nu_{\rm max, dim}$ &$\mu$Hz           &123.22 (32)    &131.31 (46)    & 89.63 (27) & 361 (86)\vspace{0.015in}\\
    Surf. cor. $a_{3, \rm bright}$&$10^{-6}$& -0.688 (73)   &-0.496 (44)    & -0.395 (42)&-0.638 (56)\vspace{0.015in}\\
    Surf. cor. $a_{-1, \rm bright}$&$10^{-6}$& -1.30 (20)   &&               &\vspace{0.015in}\\
    Surf. cor. $a_{3, \rm dim}$&$10^{-6}$& -0.418 (41)      &-0.371 (24)    & -0.399 (36)&\vspace{0.015in}\\
    Surf. cor. $a_{-1, \rm dim}$&$10^{-6}$& -0.51 (16)   &&               &\vspace{0.015in}\\ 
    Orbit $i$ \tnote{d}  & $^{\circ}$              &               &&               & 35.43 (94) \vspace{0.015in}\\
    Orbit $a$ \tnote{d}  & $\textrm{R}_{\odot}$             &               &&               & 129.6 (42)\vspace{0.015in}\\
    \hline
\end{tabular}
\begin{tablenotes}
\scriptsize
\item[a] Uncertainties are given relative to the last reported digits. 
\item[b] [Fe/H] and [$\alpha$/Fe] from APOGEE DR17 \citep{apogeedr17}. The adopted metallicity uncertainty is $0.1$ dex, since the APOGEE spectra includes companion contamination (see Sect.~\ref{sec:spec_binary_bias}).
\item[c] For KIC\,2971380, we assumed dY/dZ$\sim$1.5 \cite{Brogaard2012} and $Y_p=0.2470$ \citep{Fields2020}, since we did not have independent constraints on the modes of the dim companion.
\item[d] Including $M_{\rm bright}\sin^3 i$ and $a\sin i$ from Table~\ref{table:kic2971380}.
\end{tablenotes}
\end{threeparttable}
\end{table*}
\subsection{Inference results}
The asteroseismic inference based on radial mode frequencies converged on a single solution for all four binaries. The results of our seismic inferences are summarised in Table~\ref{table:ab_inference}.
\subsubsection{The ex situ binary, KIC\,2971380}
\begin{figure*}
    \centering
    \includegraphics[height=0.45\textheight,keepaspectratio]{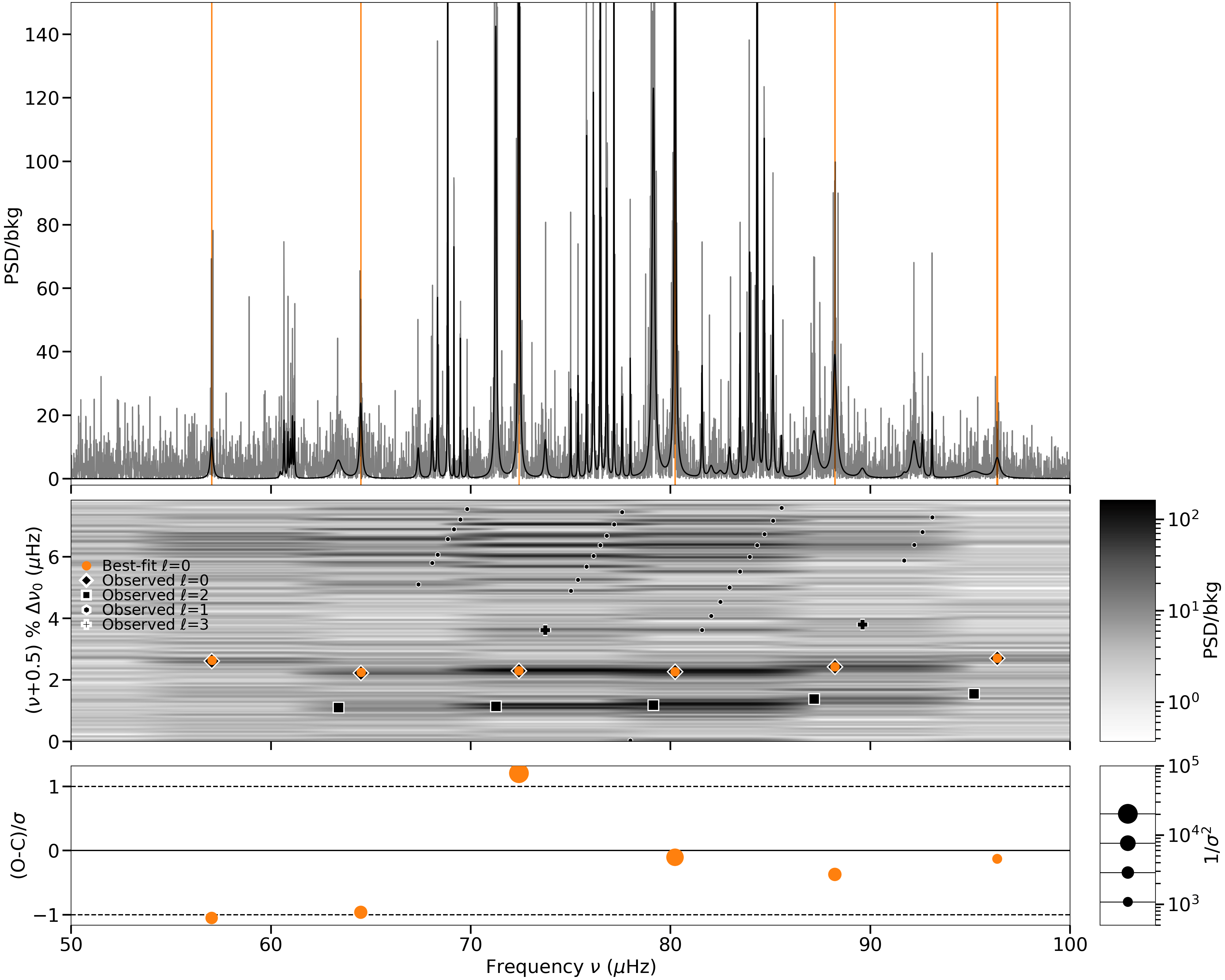}
    \caption[PSD of KIC\,2971380, including observed and theoretical frequencies.]{Top: PSD of KIC\,2971380 (grey), divided by the granulation background model. Includes best-fit peakbagging models of each oscillation mode (black). The 68$\%$ confidence interval of the radial mode frequencies is highlighted in orange. Middle: Échelle diagram (rotated), highlighting observed mode frequencies (black and white markers), and best-fit theoretical radial mode frequencies from the asteroseismic modelling inference (orange). Frequency uncertainty is included, but is often below the marker size. Bottom: Significance of the (observed - calculated) residuals for the radial mode frequencies, relative to the observation uncertainty. The marker size is scaled by the statistical weight assigned to each mode during the seismic inference. Dashed lines indicate $\pm 1\sigma$ deviation. \label{fig:psd_2971380}}
\end{figure*}

The oscillations of KIC\,2971380 are shown in Fig.~\ref{fig:psd_2971380}, together with the fit result of our seismic modelling inference.
The power spectrum of KIC\,2971380 only has detectable oscillations for the bright component.
From seismic inference, we are able to constrain the mass of the brightest star.
When combined with the RV orbital motion fit, we have inferred a stellar-model-dependent orbital inclination and semi-major axis of $35.43 \pm 0.94 ^{\circ}$ and $129.6 \pm 4.2 \textrm{R}_{\odot}$, respectively.
The radius of the brightest star relative to the orbit is therefore $4.76 \pm 0.16 \%$.
With the eccentricity of $0.264\pm 0.005$, we should not expect binary interaction to have impacted the evolution of this star yet.
As demonstrated in \citet{Montalban2021}, this system was most likely not formed in the Milky Way, but accreted later on. This is exemplified by its high Galactic eccentricity, but also by its APOGEE DR17 abundances; it has a relatively low $\alpha$ enhancement ($[\alpha/\textrm{Fe}]=0.15$) for its metallicity ($[\textrm{Fe}/\textrm{H}] = -0.77$), as well as a low $[\textrm{Al}/\textrm{Fe}]=-0.04$ and high $[\textrm{Mg}/\text{Mn}] = 0.50$ \citep[see e.g.][]{Das2020}.
Based on the binary orbital parameters, the system is similar to the thick disk benchmark dEB KIC\,10001167 studied in \citet{Thomsen2025}, with a similar orbital period, eccentricity, and semi-major axis.
It is worth mentioning that the seismically inferred age of KIC\,2971380 places it at $11.7 \pm 1.0$ Gyr, which is $1.4 \pm 1.1$ Gyr ($\sim 1.2\sigma$) older than KIC\,10001167.
This is roughly similar to the age difference found between the two systems in \citet{Montalban2021}, who used the CLES stellar evolution code \citep{Scuflaire2008}, which indicates that the result is robust against model systematics.

For our analysis of KIC\,2971380, we only included the metallicity and radial mode frequencies as observational constraints, excluding a constraint on $\nu_{\rm max}$. This way, we can make a model prediction of it using the mode frequencies, independently of the observed value. $\nu_{\rm max}$ is frequently coupled to the surface gravity, $g$, through a solar scaling relation, which is derived under the assumption that $\nu_{\rm max}$ is proportional to the acoustic cut-off frequency, $\nu_{\rm cutoff}$ \citep{Brown1991, KjeldsenBedding1995, Belkacem2011}:
\begin{align}
    \frac{\nu_{\rm max}}{\nu_{\textrm{max},\odot}} &\simeq f_{\nu_{\rm max}} \frac{g}{g_{\odot}}\left(\frac{T_{\rm eff}}{T_{\textrm{eff}, \odot}} \right)^{-1/2}.\label{eq:g_scaling}
\end{align}
We used the solar values $\nu_{\rm max, \odot}=3090 \mu$Hz, $T_{\rm eff}=5772$ K, and $g_{\odot}=27420.011$ cm/$\textrm{s}^2$ \citep{Handberg2017, IAU2015}.
The $\nu_{\rm max}$ scaling relation in this form is expected to have systematic uncertainties reflected in the unknown correction factor, $f_{\nu_{\rm max}}$ \citep[e.g.][]{Belkacem2011, Viani2017}, and recent model-dependent asteroseismic inferences of metal-poor stars appear to support this as well \citep{Yaguang2024, Huber2024, Larsen2025, Lundkvist2025, Lindsay2025}. Since we did not include $\nu_{\rm max}$ as an observational constraint for KIC\,2971380, we can predict $f_{\nu_{\rm max}}$ using the model-inferred $g$, $T_{\rm eff}$, and observed $\nu_{\rm max}$. This will be a model-dependent estimate, performed under the assumption that the individual-mode-frequency inference represents the ground truth. 
We used the model $T_{\rm eff}$ for this, because we do not have temperature observations that were derived while accounting for the companion star. As found by \citet{Yaguang2024}, model-predicted $f_{\nu_{\rm max}}$ estimates in principle depend on the model temperature scale, and thereby the mixing length. Our stellar model grid uses a fixed solar-calibrated mixing-length value, and a systematic difference between observed and model temperatures could be expected on the RGB. When using this same grid of models, the observed $T_{\rm eff}$ of the RG in the dEB KIC\,10001167 was $\sim 40$K lower than the model-inferred $T_{\rm eff}$ \citep{Thomsen2025}. When estimating $f_{\nu_{\rm max}}$ this way, we replaced the statistical uncertainty on $T_{\rm eff}$ with 80K in order to simulate the typical accuracy of spectroscopy \citep{Bruntt2010}.

\begin{figure}
    \centering
    \includegraphics[width=\columnwidth]{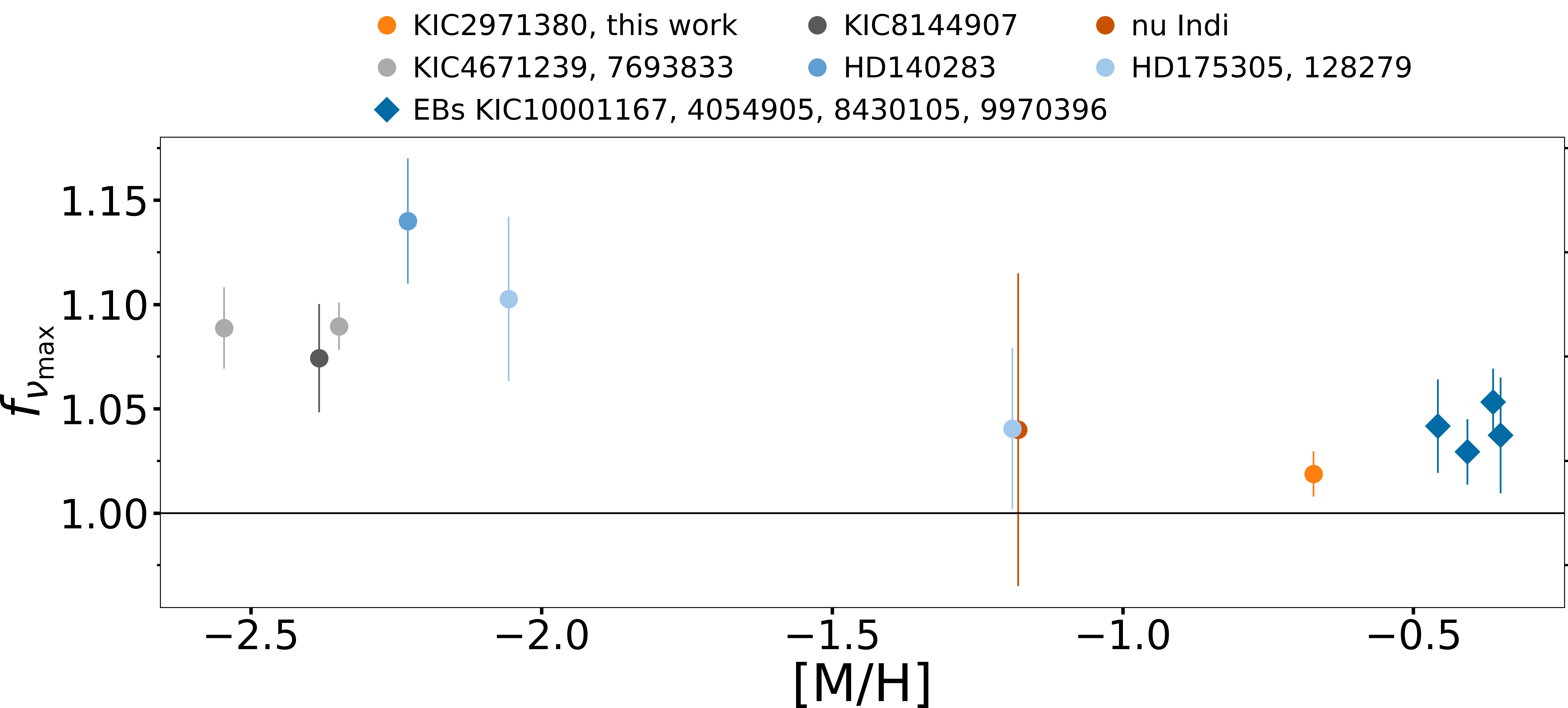}
    \caption{$f_{\nu_{\rm max}}$ estimates for low-metallicity stars (see text for details). }
    \label{fig:f_numax}
\end{figure}

With this we derived $f_{\nu_{\rm max}} = 1.019 \pm 0.011$, which is of low significance but indicative of the need for a correction larger than one for metal-poor stars. Particularly if $f_{\nu_{\rm max}}$ is increasing with decreasing metallicity, this would be consistent with the larger corrections inferred from individual mode frequencies for the most metal-poor oscillators (KIC\,8144907 \citep{Huber2024}, KIC\,4671239 \& 7693833 \citep{Larsen2025}, HD140283 \citep{Lundkvist2025}, HD175305 \& 128279 \citep{Lindsay2025}, and $\nu$ Indi \citep{Chaplin2020}). We can further compare with the best-studied low-metallicity dEBs in \emph{Kepler} that also contain a solar-like oscillator, KIC\,10001167, 4054905, 8430105, and 9970396 \citep{Thomsen2025, Brogaard2022, Thomsen2022, Brogaard2018}. For these dEBs, the surface gravity measurement is dependent only on the orbital and eclipse modelling, and therefore fully independent of stellar evolutionary models. We show the comparison in Fig.~\ref{fig:f_numax}. The dEBs further support the need for a correction larger than unity for low-metallicity stars.

From the spectroscopy (see Fig.~\ref{fig:2971380_bf}), the ratio of the line profile integrals of the binary is roughly $25\%$.
While the exact value is dependent on the assumed spectroscopic template, the true luminosity ratio of the binary is likely within the same order of magnitude.
Together with the RV mass ratio, we have used this to estimate the properties of the dimmer companion, assuming $L_{\rm dim}/L_{\rm bright} \sim 0.25 \pm 0.05$ and that the two stellar ages agree to at least $1\%$. From stellar inference, we estimate that the companion is compatible with a star at the base of the RGB, having $R \sim 2.8 \textrm{R}_{\odot}$ and $\nu_{\rm max} \sim 361 \pm 86 \mu$Hz. This puts it at or above the Nyquist frequency of \emph{Kepler} long-cadence data.
Unfortunately, with an apparent magnitude of $G=12.6$, the system is far too dim for the TESS photometry to have photometric precision sufficient to measure oscillations from either star.
With a future pointing of the PLATO space mission \citep{Rauer2025}, or with the Earth 2.0 telescope \citep{Earth2.0_2022}, this system should be considered a strong target of interest for detection of the companion. With the upcoming Gaia Data Release 4, a potential astrometric orbit fit would allow for a fundamental measurement of the mass of the brightest star, providing the first accreted star with solar-like oscillations and an absolute mass measurement. This would make KIC\,2971380 an excellent benchmark system for Galactic archaeology and for studying the early merger history, together with the old thick disk dEB KIC\,10001167 \citep{Thomsen2025}.
\subsubsection{The AB(C)s KIC\,6501237, 10592924, and 11299484}
For KIC\,6501237 and 10592924, Fig.~\ref{fig:corner_6501237} and \ref{fig:corner_10592924} in the appendix show the posterior distributions of the dim and bright star. The posterior distribution of KIC\,11299484 behaves similarly.
In the appendix, Fig.~\ref{fig:psd_6501237A}, \ref{fig:psd_6501237B}, \ref{fig:psd_10592924A}, \ref{fig:psd_10592924B},  \ref{fig:psd_11299484A}, and \ref{fig:psd_11299484B} show the power spectra, observed frequencies, and theoretical frequencies for the three systems. Note that we excluded one radial mode from the brighter star of KIC\,6501237, due to its high uncertainty and very low peakbagged amplitude. Given the large uncertainty, it would not have affected the inference results.

When we include only a cubic term to the surface correction, the best-fitting theoretical frequencies of KIC\,6501237 deviate significantly from the observed. For the brighter star, most modes were $1-5\sigma$ different, while for the dimmer component, all modes deviated by 1-1.5$\sigma$.
We verified that this discrepancy remains when fitting the two stars separately. When we included an additional inverse term to the \citet{ballgizon2014} surface correction, we recovered a good fit of the modes. However, the inverse term of the brighter star is an order of magnitude larger than the cubic one, which is not expected to be correct \citep{ballgizon2014}. This suggests that the discrepancy stems from a systematic difference between the stellar models and the two stars.
Despite this, the seismic inference of KIC\,6501237 can clearly be reproduced with a coeval model. The stellar ages agree far better ($0.4\%$) than expected from the Gaussian prior ($5\%$). If the dimmer set of oscillations comes from the dimmer optically resolved star located at $\sim 3"$ on-sky separation, the two stars must either be in a very long period orbit, or part of the same stellar association. Based on the predictions of \citet{Miglio2014}, all optically resolved ABCs in \emph{Kepler} should in principle be false positives (chance alignments), and the confirmation of such a system is therefore notable.

For KIC\,10592924 and KIC\,11299484, we find a reasonable agreement between theoretical and observed oscillation mode frequencies when using just the cubic correction. The fit quality of the brighter star in KIC\,10592924 was marginally better when including an inverse term. As with KIC\,6501237, the high fit quality supports that the stars are well described as coeval.

\begin{figure}
    \centering
    \includegraphics[width=\columnwidth]{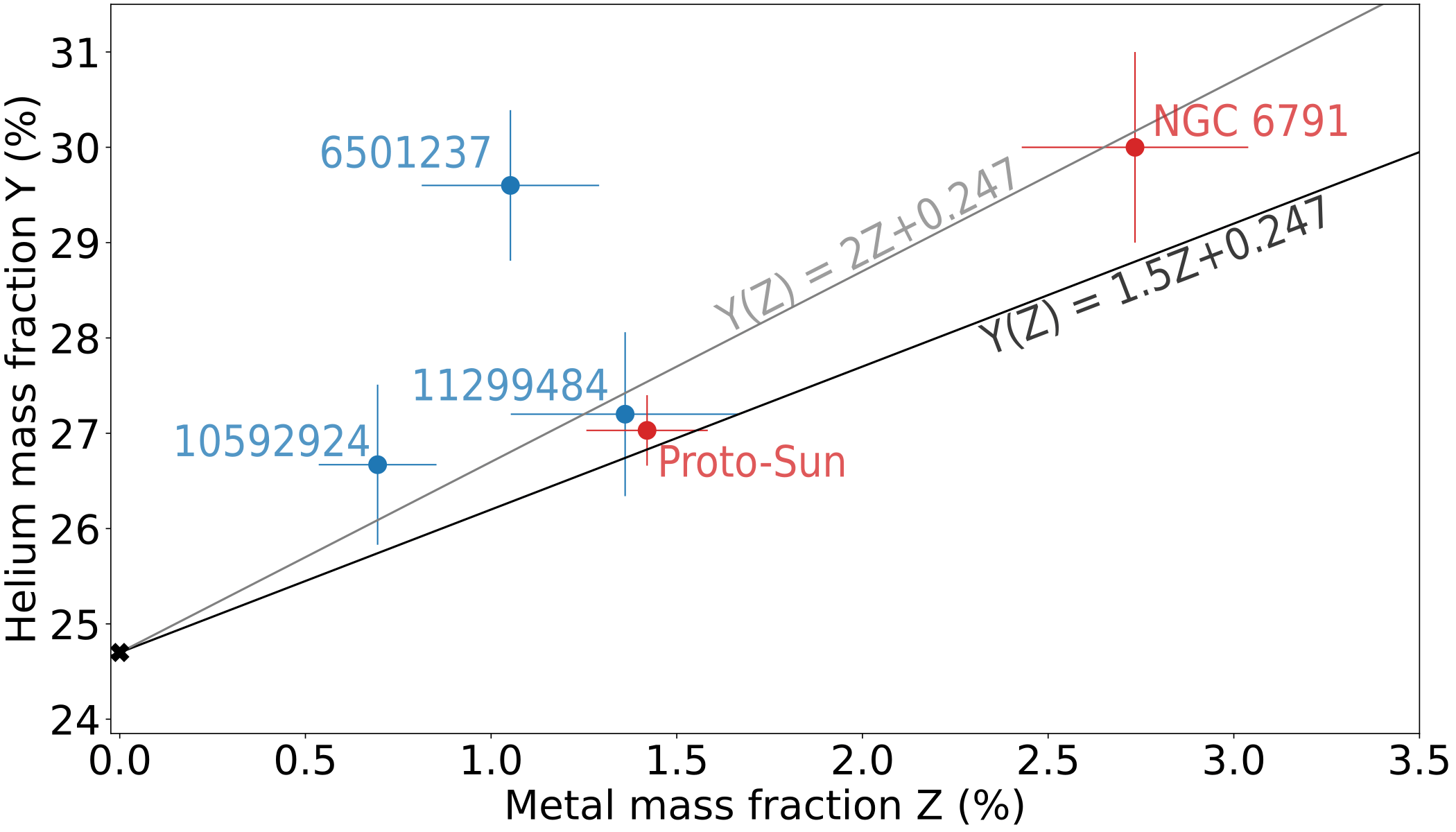}
    \caption{Helium mass fraction inferred for the ABs KIC\,6501237, 10592924, and 11299484, compared to the proto-Sun \citep{Asplund2009} and NGC\,6791 \citep{Brogaard2012}. We have also assumed the \citet{Asplund2009} present-day solar composition.}
    \label{fig:helium}
\end{figure}

For all three, it appears to be possible to infer the initial helium content, at $3\%$ precision for KIC\,10592924 and 11299484, and at $4\%$ precision for KIC\,6501237. We have illustrated this in the corner plots in the appendix (Fig.~\ref{fig:corner_6501237} and \ref{fig:corner_10592924}). In comparison, we could infer helium to $5\%$ precision when studying only the brightest oscillator of an AB. \citet{Mckeever2019} similarly performed seismic modelling of RGB stars in NGC\,6791, and showed that inferring a precise helium mass fraction from a single RGB star can be challenging. 
Since the stars are on the RGB, the treatment of diffusion and gravitational settling is not expected to have a significant impact on the inferred initial helium values, which is generally a major challenge for main-sequence seismology \citep[e.g. ][]{VermaAguirre2019,Nsamba2021}. This makes RGB+RGB ABs potential helium calibration systems. In Fig.~\ref{fig:helium}, we compare the inferred helium values to the proto-Sun \citep{Asplund2009}, NGC\,6791 \citep{Brogaard2012}, and Big Bang nucleosynthesis \citep{Fields2020}. We remark that including an inverse surface term for the brighter star in KIC\,10592924 (one-term or two-term) increased the inferred helium value at $\sim 0.9\sigma$ significance.
In addition, the derived helium value of KIC\,6501237 is $\sim 2\sigma$ larger than what is expected for the system's metallicity (for illustration, see Fig.~\ref{fig:helium}). Given the need for a large inverse term to the surface effect for KIC\,6501237, we tentatively conclude that we have over-fitted its observed frequencies with the additional surface term.

The inference of a precise helium value relied on a $\nu_{\rm max}$ constraint, using the asteroseismic scaling relation (eq.~\eqref{eq:g_scaling}).
This might be avoided if we had observational constraints on the stellar luminosities; however, this is beyond the current scope of this work.
Therefore, we note that these first helium measurements, precise but systematically uncertain, should serve as an encouraging motivation to study ABs. In a future study, we shall return to these systems with a detailed modelling effort.

The dimmer companions in KIC\,6501237 and 10592924 seem to show peculiar dipole modes (Fig.~\ref{fig:psd_6501237B}, ~\ref{fig:psd_10592924B}). Particularly for KIC\,10592924, the dipole modes appear depressed compared to the other mode amplitudes. This can be an indication of magnetic activity \citep[e.g.][]{Fuller2015, Stello2016}.
\section{Conclusion}\label{sec:conclusion}
Binary stars containing two oscillating RGs offer particularly powerful constraints on models of stellar evolution, if the oscillations of the two stars can be disentangled. With high-resolution spectroscopy, we have detected RV variability in 12 candidate binary systems, consistent with them being single-lined or double-lined binaries. Eight of them are also detected as ABs, with three having indications of two overlapping sets of oscillations, four having separate oscillation envelopes, and one (KIC\,5015647) showing a very complex power spectrum indicative of a non-solar-like oscillating companion. An additional 5-6 systems have spectroscopic line shapes that are atypical of RGs, but with only a single observation with FIES, we have not been able to explore variability in their spectra. By cross-matching our observations with Gaia DR3 and APOGEE DR17, we find an additional 9 likely binaries (6 ABCs), most of which have tentative but supporting indications of binarity across several sources of measurement. Additionally, we confirmed the high-mass evolved oscillator KIC\,5978324 \citep{Crawford2024, Crawford2025} as RV variable based on three spectra obtained over 406 days, with $\textrm{d}v/\textrm{d}t = -2.85 \pm 0.05 $kms/year.

With eight spectra of a binary system of extragalactic origin, KIC\,2971380, we were able to constrain the orbital motion, and confirm its mass ratio as consistent with unity to 1\% precision. This suggests that the companion is also evolved past the main sequence. With asteroseismic inference of the brighter star, we made predictions for the properties of the companion and the orbit, estimating $\nu_{\rm max, dim} \sim 361\pm86\mu$Hz, orbit inclination $i\sim 35.4\pm 0.9^{\circ}$, and semi-major-axis $130 \pm 4 \textrm{R}_{\odot}$. This makes it a prospective target for detection of the companion's oscillations with either PLATO \citep{Rauer2025} or the Earth 2.0 Telescope \citep{Earth2.0_2022}.
We derived an age of $11.7\pm 1.0$Gyr, consistent with it being $\sim 1.4 \pm 1.1$Gyr older than the thick disk dEB KIC\,10001167 \citep{Thomsen2025}. Astrometry from the upcoming Gaia DR4 \citep{GaiaMission2016} might allow us to obtain a precise constraint on the stellar masses.

We performed a double-seismic inference, utilising the oscillation frequencies of both stars, for two ABCs, namely KIC\,6501237 and 11299484, and the confirmed AB KIC\,10592924. All three targets contain two RGB oscillators. With our inference, we confirmed that the oscillations of their two stellar components are well-modelled as coeval. We obtained precise constraints on their masses ($\sim 1.5\%$), mass ratio ($\sim 1\%$), and the age of the binary ($ 8-10\%$), while inferring their initial helium content ($\sim 3\%$). Using only the brightest oscillator, the helium could be constrained to $\sim 5\%$ precision. For KIC\,10592924 and 11299484, we derived helium values consistent with typically assumed linear helium-metallicity relations. Particularly KIC\,11299484 agrees well with proto-solar predictions \citep{Asplund2009}.

Helium inferred from RGs is affected by different systematic uncertainties from main-sequence stars. A major challenge for the former is the treatment of diffusion and gravitational settling \citep[e.g.][]{VermaAguirre2019, Nsamba2021}. The uncertainty in the stellar helium content contributes a large systematic uncertainty to stellar parameters, including the age \citep[see e.g. Fig.~43 of ][]{Lebreton2014}. This impacts several fields relying on accurate stellar model parameters and ages, including Galactic archaeology and exoplanetary studies. With a precise helium inference, we demonstrate that ABs with RGs can serve as helium calibration laboratories.
\begin{acknowledgements}\\
We wish to thank the anonymous referee for their thorough review, which led to a significant improvement of the manuscript.

We are grateful to Josefina Montalbán for a large number of target suggestions for ABCs to observe, and for helpful comments on the manuscript draft. 

JST, AMazzi acknowledge support from Bologna University, "MUR FARE Grant Duets CUP J33C21000410001".

AMiglio, MMatteuzzi acknowledge support from the ERC Consolidator Grant funding scheme (project ASTEROCHRONOMETRY, \url{https://www.asterochronometry.eu}, G.A. n. 772293).

J.Y.C acknowledges support from the Klaus Tschira Foundation and funding from the ERC Consolidator Grant DipolarSound (grant agreement \# 101000296).

Based on observations made with the Nordic Optical Telescope, owned in collaboration by the University of Turku and Aarhus University, and operated jointly by Aarhus University, the University of Turku and the University of Oslo, representing Denmark, Finland and Norway, the University of Iceland and Stockholm University at the Observatorio del Roque de los Muchachos, La Palma, Spain, of the Instituto de Astrofisica de Canarias.
\end{acknowledgements}
\bibliographystyle{aa}
\bibliography{references}

\label{LastPage}
\clearpage

\begin{appendix}
\nolinenumbers
\twocolumn
\section{Additional tables and figures}
\FloatBarrier
\begin{figure}[h!]
    \centering
    \includegraphics[width=\columnwidth]{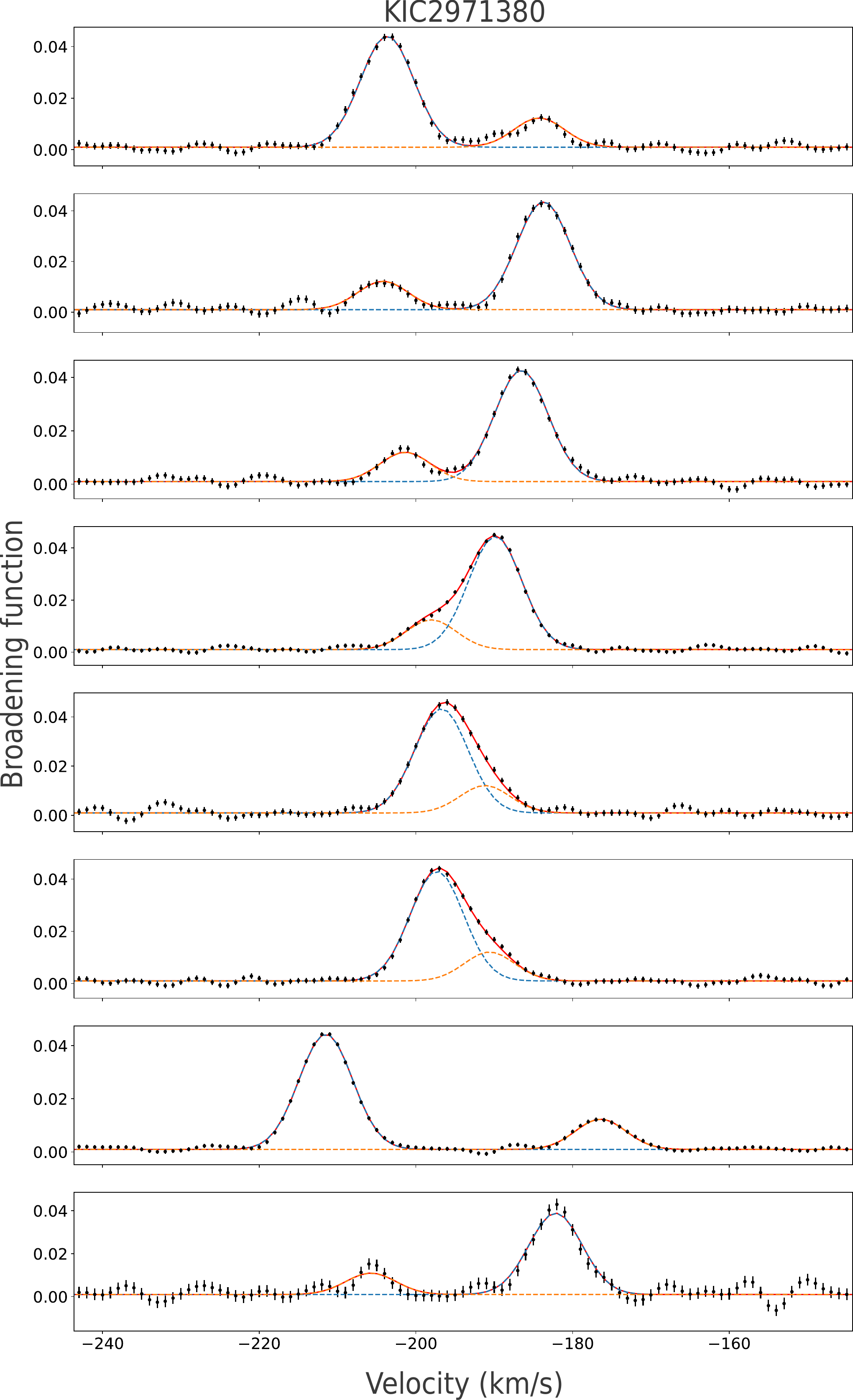}
    \caption{Simultaneous double-Gaussian fit to the BFs of all spectra of KIC\,2971380. The combined model is shown in the red full line, while orange and blue dashed lines are the two components.}
    \label{fig:2971380_bf}
\end{figure}
\begin{table}
    \centering
\caption[Average asteroseismic parameters for four binaries.]{\footnotesize Average asteroseismic parameters for four binaries.}\label{table:ab_obs_seis}
\begin{threeparttable}
    \scriptsize
\begin{tabular}{lccccc}
    KIC\tnote{a} & & 6501237 & 10592924 & 11299484 & 2971380 \\
    \hline
    \multicolumn{3}{l}{Asymp. fit (Bright)}  & & & \\
    \hline
    $\nu_{\rm max}$\tnote{b} & $\mu$Hz      & 62.86 (19)    & 52.52 (24)     & 48.27 (23) & 77.63 (26) \\
    $\Delta\nu_0$         & $\mu$Hz         & 6.1538 (42)   & 5.3827 (42)   & 4.9899 (48)  & 7.8457 (45) \\
    $\Delta\nu_{\rm obs}$ & $\mu$Hz         & 6.1822 (81)   & 5.402 (11)     & 5.056 (12) & 7.920 (14) \\
    $\delta\nu_{\rm obs, 02}$ & $\mu$Hz     & 0.831 (29)    & 0.743 (39)     & 0.673 (39) & 1.115 (49)\\
    $\epsilon_{\rm obs}$ &                  & 0.075 (14)    & 0.084 (20)     & 0.062 (22) & 0.131 (19) \\
    $\alpha_{\rm obs}$ &                    & 0.0145 (16)   & 0.0168 (24)    & 0.0168 (33) & 0.0154 (27)\\
    $\sigma_{\rm jitter}$ & $\mu$Hz         & 0.046 (13)    & 0.065 (18)    & 0.059 (19)  & 0.068 (22)\\
    \hline
    \multicolumn{3}{l}{Asymp. fit (Dim)}  &  & & \\
    \hline
    $\nu_{\rm max}$\tnote{b} & $\mu$Hz      & 123.18 (41)   & 130.56 (82)  & 89.82 (41) &   \\
    $\Delta\nu_0$         & $\mu$Hz         & 10.4818 (77)  & 11.0416 (76) & 8.1095 (86) & \\
    $\Delta\nu_{\rm obs}$ & $\mu$Hz         & 10.538(20)    & 11.059 (15)  & 8.154 (25) & \\
    $\delta\nu_{\rm obs, 02}$ & $\mu$Hz     & 1.339 (52)    & 1.436 (38)   & 1.084 (44) & \\
    $\epsilon_{\rm obs}$ &                  & 0.180 (22)    & 0.246 (16)   & 0.159 (34) & \\
    $\alpha_{\rm obs}$ &                    & 0.0092 (33)   & 0.0000 (22)  & 0.0183 (63) & \\
    $\sigma_{\rm jitter}$ & $\mu$Hz     & 0.065 (24)    & 0.049 (18)    & 0.056 (28) & \\
    \hline
\end{tabular}
\begin{tablenotes}
\scriptsize
\item[a] Uncertainties are given relative to the last reported digits.
\item[b] From the background model fit. See text for details.
\end{tablenotes}
\end{threeparttable}
\end{table}
\newpage
\FloatBarrier
\onecolumn
\begin{table*}[h!]       
    \centering
    \caption[Summary of spectroscopic monitoring of candidate binary systems.]{\label{table:spec_abs}RV variability summary for targets monitored with FIES over multiple nights.}
    \begin{threeparttable}
    \begin{tabular}{lccccccccc}
        \hline\hline KIC    & $\frac{\Delta\sigma}{\textrm{Err}}$\tnote{a,b} & $v$\tnote{c}& $\Delta v$\tnote{b,c} & $\frac{\textrm{d}v}{\textrm{d}t}$ (kms/yr)\tnote{c,d} & $\Delta$BIC\tnote{e} & baseline (days) & \#\tnote{f} & model\tnote{g} & source\tnote{h}\\\hline
        \multicolumn{6}{l}{Significant variability} &&&\\ 
        \hline
        2971380     &7.1   &35.0      & 50.03 (29)    & 40.18 (68)    & -3530 & 189   & 7 & SB2    & 1\\
        4260884     &3.8   &3.6       & 3.68 (34)     & 7.84 (68)     & -101  & 180   & 4 & SB2    & 4\\
        5300269     &1.7   &2.9   & 4.1 (10)      & -7.8 (18)     & -18   & 201   & 3 & SB2        & 1,2\\
        7697607     &2.1    &9.8   & 0.35 (15)    & -0.37 (12)   & -8   & 343   & 3 & SB2           & 1,2\\
        8479182     &0.8    &11.8  & 0.38 (16)    & -0.33 (15)   & -3   & 351   & 5 & SB2           & 1,3\\
        8479383     &8.7     &10.8  & 8.19 (18)     & 10.99 (24)    & -2081 & 275   & 4 & SB2        & 1,4\\
        8649099     &3.3   &9.8   & 0.73 (15)    & 0.92 (18)    & -24   & 286   & 5 & SB2          & 1\\
        10592924    &0.1  & 4.8       & 9.4 (10)      & 25.1 (27)     & -87   & 137   & 2 & SB2   & 1,2\\
        4346953     &0.03    & -0.8      & 2.77 (14)    & -7.08 (37)    & -372 & 143   & 2 & SB1     & 4\\
        4659808     &1.0    & 19.5      & 2.23 (15)    & -5.59 (37)    & -232  & 146   & 2 & SB1    & 4\\
        5015647     &0.4    & -32.0     & 7.23 (14)    & -18.75 (37)   & -2532 & 141   & 2 & SB1    & 2\\
        5978324     &0.4    & 18.7      & 3.20 (14)    & -2.85 (11)   & -634 & 406   & 3 & SB1      & 3\\
        \hline
        \multicolumn{6}{l}{Tentative indication of variability}&&&\\ 
        \hline
        11299484    &1.1   &5.66   & 0.45 (25)     & 0.37 (21)     & -2  & 255   & 3 & SB2         & 1,2,4\\
        4637699     &0.6    & -97.0         & 0.28 (23)     & 0.55 (47)     & 0  & 146   & 2 & SB1  & 1,3\\
        6501237\tnote{i}     &2.7    & -7.3          & 0.23 (15)    & -0.49 (32)    & -2  & 170   & 2 & SB1 & 1,2\\
        8181509     & 0.01   & -31.2     & 0.27 (14)    & 0.24 (13)    & -3   & 413   & 2 & SB1              & 1,3\\
        9392647\tnote{j}     & 0.3  & -44.7         & 0.27 (18)     & 2.1 (14)      & -1  & 42    & 2 & SB1 & 2,3\\
        9392650\tnote{j}     & 0.5   & -83.5         & 0.39 (18)     & 3.4 (16)      & -4  & 42    & 2 & SB1 & 2,3\\
        9412408     & 0.2   & 10.3  & 0.13 (14)    & -0.18 (20)    & -0  & 259   & 2 & SB1                   & 1,4\\
        \hline
        \multicolumn{6}{l}{Insignificant variability}&&&\\\hline 

        6369592\tnote{k}     &1.8    &21.12  & 0.65 (23)     & -0.11 (17)    & 2   & 413   & 4 & SB2                 & 3\\
        2442606     &0.2    & 16.5      & 0.04 (14)    & 0.22 (91)     & 1   & 58    & 2 & SB1              & 3\\
        4157282     &1.0    & -19.6     & 0.01 (15)    & -0.03 (46)    & 1   & 119   & 2 & SB1              & 1\\
        4159205     &0.3    & -62.8     & 0.02 (14)    & 0.01 (17)    & 1   & 259   & 2 & SB1               & 1\\
        5024272     &0.2    & 4.5       & 0.06 (14)    & -0.23 (56)    & 1   & 94    & 2 & SB1              & 1\\
        5202286     &0.7    & 16.6      & 0.08 (15)    & -0.09 (18)    & 0  & 307   & 2 & SB1               & 1\\
        5353108     &0.2    & -50.5     & 0.05 (14)    & -0.15 (41)    & 1   & 126   & 2 & SB1              & 1\\
        5630997     &0.1    & 9.3       & 0.03 (18)     & 0.05 (29)     & 1   & 227   & 2 & SB1             & 1\\
        6206407     &0.3    & -47.7     & 0.02 (14)    & 0.04 (23)     & 1   & 227   & 2 & SB1              & 3,4\\
        6208846     &0.2   & -7.8      & 0.05 (14)    & 0.12 (36)     & 1   & 143   & 2 & SB1              & 1\\
        6221548     &0.1   & -41.2     & 0.04 (14)    & 0.11 (40)     & 1   & 133   & 2 & SB1              & 4\\
        6526377     &0.2   & -6.0      & 0.01 (14)    & 0.02 (20)     & 1   & 262   & 2 & SB1              & 2,4\\
        8708536     &0.1   & -18.0     & 0.03 (14)    & -0.04 (20)    & 1   & 263   & 2 & SB1              & 2\\
        9205419     &0.6   & -4.7      & 0.07 (14)    & 0.04 (13)    & 2   & 276   & 3 & SB1               & 1,2\\
        9775927     &0.1   & -10.0     & 0.01 (14)    & 0.01 (20)     & 1   & 259   & 2 & SB1              & 3,4\\
        10464837    &0.3   &-22.4      &0.05 (14)     & 0.05 (15)    & 1  & 297   & 2 & SB1                & 1\\
        11352644    &0.8    & -52.0         & 0.05 (15)    & -0.03 (15)   & 1   & 300   & 2 & SB1            & 1\\
        11136690    &0.8    & -59.3     & 0.04 (14)    & 0.01 (16)    & 1   & 255   & 3 & SB1               & 1\\
        \hline
    \end{tabular}
    \begin{tablenotes}
        \scriptsize
        \item[a] Difference in width of a Gaussian model fit, scaled by uncertainty, assuming only one spectral line.
        \item[b] Most significant difference between two spectra.
        \item[c] For SB1 models, this is measured in RV. For SB2 models, this is based on $\Delta v= RV_2-RV_1$.  Velocity units are in km/s unless specified.
        \item[d] Slope of a linear fit.
        \item[e] Difference between BIC of a linear and a constant fit. Negative favors the linear.
        \item[f] Number of observed nights.
        \item[g] Double-lined (called SB2) model fits were only attempted on visually bimodal spectra. If an SB2 model was used, the single-lined (SB1) results are not reported (except for $\Delta\sigma$).
        \item[h] Source for the candidate binary. 1: Candidate photometric AB (Mazzi et al. (in prep.)). 2: ABCs and poorly fitting single-star background fits from Josefina Montalbán (priv. comm.). 3: ABCs and strange oscillators flagged by manual inspection of KEPSEISMIC power spectra. 4: ABCs, candidate binaries with solar-like oscillators, and peculiar oscillators presented in \citet{Colman2017, Bell2019, Gaulme2020, Schonhut-Stasik2020}.
        \item[i] KIC\,6501237 has a nearby star with similar RV, and seeing conditions were significantly different for the two nights.
        \item[j] KIC\,9392647 and 9392650 are very close on the sky, and was likely selected as candidate ABs because of that. While their RVs are different, their proximity might have affected the observation quality.
        \item[k] Binary solution ruled out, see Appendix~\ref{sec:spec_interesting_targets}.
    \end{tablenotes}
    \end{threeparttable}
\end{table*}
\begin{table*}       
    \centering
    \footnotesize
    \caption{\label{table:spec_1spec}Width $\sigma$ and skew $\alpha$ of spectroscopic line profile for all stellar systems observed with FIES.}
    \begin{threeparttable}
    \begin{tabular}{llcccccccc}
        \hline\hline &KIC    & $\sigma$ (km/s) & $\alpha$ & RUWE\tnote{a} & Var/DR3 (\#visit)\tnote{a} & NSS\tnote{a} & RV amp\tnote{a} & VSCATTER (\#visit)\tnote{b} & source\tnote{c}\\\hline
        \parbox[t]{2mm}{\multirow{23}{*}{\rotatebox[origin=c]{90}{Only one night\tnote{d}}}} 
&1720241  & 3.186 (27)  & 1.02 (12) & 1.05 & False (29) & 0 & 14.25 & 0.06 (2)  & 3\\
&2304765  & 3.645 (16)  & 1.053 (61) & 0.91 & False (26) & 0 & 4.41 & 0.02 (3)  & 1\\
&2447476  & 3.187 (18)  & 0.872 (92) & 1.05 & False (26) & 0 & 5.74 & 0.04 (2)  & 1\\
&3660820  & 3.394 (18)  & 1.130 (70) & 0.78 & False (19) & 0 & 3.38 & 0.03 (3)  & 3,4\\
&3855003  & 3.266 (25)  & 1.185 (99) & 0.98 & False (20) & 0 & 5.85 & 0.13 (2)  & 2\\
&3860550  & 5.403 (31)  & 0.38 (32)  & 0.95 & False (24) & 0 & 5.88 & 0.02 (3)  & 1\\
&3955502  & 5.814 (44)  & -0.33 (55) & 0.88 & False (21) & 0 & 7.80 & 0.02 (2)  & 1\\
&4058578  & 3.323 (27)  & 0.2 (23)  & 0.98 & False (22) & 0 & 8.90 &  & 2\\
&4282111  & 3.791 (33)  & 0.65 (21) & 1.19 & False (16) & 0 & 11.83 & 0.12 (3) & 1 \\
&4768054  & 3.402 (32)  & 0.89 (16)  & 0.97 & False (18) & 0 & 6.10 &  & 2\\
&5647777  & 3.802 (29)  & 1.06 (11)  & 1.01 & False (15) & 0 & 14.96 & 0.07 (2) & 1\\
&5700488  & 3.550 (26)  & 1.551 (86)  & 1.07 & False (16) & 0 & 3.42 & & 1\\
&5943448  & 3.205 (19)  & 1.191 (78)  & 0.87 & False (22) & 0 & 8.44 & & 2\\
&6210603  & 3.501 (9)  & 1.064 (36) & 0.84 & False (8) & 0 & 1.46 &  & 2\\
&6706199  & 3.626 (31)  & 1.54 (10)  & 1.12 & False (21) & 0 & 5.48 & 0.01 (2) & 1\\
&6707691  & 3.296 (37)  & 1.01 (16)  & 1.93 & False (19) & 0 & 7.46 & & 2,4\\
&6841636  & 3.739 (22)  & 0.922 (93) & 1.08 & False (17) & 0 &  &   & 1 \\
&7287493  & 3.409 (13)  & 1.069 (53) & 1.07 & False (20) & 0 & 4.10 & & 2 \\
&8295007  & 3.263 (15)  & 1.046 (65) & 1.12 & False (19) & 0 & 8.75 & 0.24 (2) & 2 \\
&10841730  & 3.649 (16)  & 0.924 (70) & 47.53 & True (7) & 0 & 8.24 & 0.06 (2) & 2 \\
&11465942  & 3.375 (13)  & 1.194 (49) & 1.69 & True (19) & 0 & 10.74 & & 2,4 \\
&11656670  & 3.481 (22)  & 1.242 (77) &  & True (20) & 0 & 4.58 & & 2  \\
&12117790  & 3.135 (19)  & 0.86 (10)  & 4.04 &  & 0 &  & 0.06 (2) & 1,4 \\
        \hline
        \multicolumn{3}{l}{Single-lined RV variable\tnote{d}}\\ 
&4346953  & 3.284 (13)  & 1.044 (52) & 1.37 & True (18) & 2 & 11.92 & 0.12 (5) & 4\\
&4659808  & 4.676 (20)  & 0.389 (40) & 1.35 & True (20) & 2 & 10.11 & 0.10 (4) & 4\\
&5015647  & 3.484 (13)  & 1.033 (62) & 10.31 & True (22) & 2 & 15.63 & 1.85 (4) & 2\\
&5978324  & 3.825 (11)  & 0.889 (44) & 2.54 & True (20) & 2 & 9.02 &  & 3\\
        \hline
        \parbox[t]{2mm}{\multirow{10}{*}{\rotatebox[origin=c]{90}{Double-lined\tnote{d}}}}
&2971380  & 3.662 (23)  & 0.041 (99) & 1.09 & True (22) & 0 & 26.68 & 0.51 (2)  & 1 \\
&4260884  & 4.107 (13)  & -0.572 (36) & 13.04 & & 0 &  & 0.02 (2) & 4\\
&5300269  & 4.267 (22)  & -0.116 (45) & 0.94 & False (22) & 0 & 7.92 &  & 1,2\\
&6369592  & 3.329 (13)  & 0.827 (49) & 4.81 & False (17) & 0 & 11.50 &  & 3\\
&7697607  & 6.786 (23)  & -5.127 (81)  & 0.95 & False (29) & 0 & 9.46 & & 1,2\\
&8479182  & 8.153 (38)  & -2.62 (12) & 0.90 & False (18) & 0 & 3.36 & 0.10 (5) &1,3 \\
&8479383  & 4.146 (11)  & 2.129 (37) & 1.17 & True (16) & 0 & 11.13 & 0.17 (5) &1,4\\
&8649099  & 6.905 (88)  & -3.1 (17) & 0.93 & False (16) & 0 & 3.34 & 0.35 (2)  &1\\
&10592924  & 4.272 (25)  & 0.632 (69) & 1.11 & True (15) & 0 & 12.77 &  &1,2\\
&11299484  & 4.516 (20)  & 1.459 (58) & 1.09 & False (14) & 0 & 11.89 & 0.60 (3) &1,2,4\\
        \hline
        \parbox[t]{2mm}{\multirow{23}{*}{\rotatebox[origin=c]{90}{Single-lined non-detection\tnote{d}}}}
&2442606  & 3.387 (12)  & 1.019 (65)  & 0.94 & False (15) & 0 & 2.46 & 0.06 (5) & 3\\
&4157282  & 3.391 (22)  & 1.064 (95) & 0.91 & False (15) & 0 & 1.91 & 0.07 (3)  & 1 \\
&4159205  & 3.430 (8)  & 0.984 (36)  & 1.07 & False (21) & 0 & 14.12 & 0.12 (5) & 1 \\
&4637699  & 4.070 (40)  & -0.23 (17)  & 1.09 & False (22) & 0 &  &  &  1,3 \\
&5024272  & 3.547 (13)  & 0.94 (10)  & 0.94 & False (16) & 0 & 3.92 & 0.13 (2) & 1\\
&5202286  & 3.352 (23)  & 1.090 (16) & 0.98 & False (23) & 0 & 2.90 & 0.03 (3) & 1\\
&5353108  & 3.508 (6)  & 1.107 (22) & 0.86 & False (16) & 0 & 0.83 & 0.04 (2) & 1\\
&5630997  & 3.480 (46)  & 1.13 (16)  & 1.03 & False (19) & 0 & 1.58 & 0.05 (4) &1\\
&6206407  & 3.335 (13)  & 1.053 (57) & 34.04 & & 0 &  &  & 3,4 \\
&6208846  & 3.425 (9)  & 1.085 (34)  & 0.95 & False (18) & 0 & 1.15 & 0.05 (2) & 1 \\
&6221548  & 3.801 (13)  & 0.931 (55)  & 0.94 & False (18) & 0 & 2.88 & 0.05 (2) & 4 \\
&6501237  & 3.701 (19)  & 1.088 (70) & 1.03 & False (18) & 0 & 6.41 &  & 1,2\\
&6526377  & 3.414 (12)  & 1.154 (46) & 0.89 & False (6) & 0 & 1.22 & 0.02 (2) &  2,4\\
&8181509  & 3.530 (7)  & 1.085 (27)  & 3.10 & True (15) & 2 & 5.35 & 0.22 (2) & 1,3 \\
&8708536  & 3.324 (7)  & 1.081 (29)  & 0.87 & False (18) & 0 & 2.61 &  & 2\\
&9205419  & 3.389 (10)  & 0.761 (40) & 1.06 & False (21) & 0 & 16.74 & 0.07 (3) & 1,2 \\
&9392647  & 3.153 (20)  & 0.713 (78) & 1.05 & False (25) & 0 &  &  & 2,3 \\
&9392650  & 3.420 (46)  & 1.080 (16) & 0.88 & False (30) & 0 & 3.95 & 0.09 (5) & 2,3 \\
&9412408  & 3.397 (13)  & 1.026 (65) & 0.94 & False (15) & 0 & 1.38 & & 1,4 \\
&9775927  & 3.536 (8)  & 1.080 (31) & 1.01 & False (11) & 0 & 2.81 &  & 3,4 \\
&10464837  & 3.481 (10)  & 1.044 (42) & 1.02 & False (27) & 0 &  &   & 1\\
&11136690  & 3.765 (14)  & 0.959 (57) & 0.93 & False (22) & 0 & 13.00 & 0.52 (3) & 1 \\
&11352644  & 3.519 (15)  & 0.951 (64) & 1.09 & False (15) & 0 &  & 0.61 (3) & 1 \\
        \hline
    \end{tabular}
    \begin{tablenotes}
        \scriptsize
        \item[a] Gaia DR3 \citep{GaiaDR3, Katz2023, GaiaDR3NSS}. RUWE=Renormalised Unit Weight Error, $>> 1$ is indicative of astrometric excess signal. Var/DR3 is a significant indicator of RV variability, requiring rv\_chisq\_pvalue $ < 0.01$ and rv\_renormalised\_gof $ > 4$ \citep{Katz2023}, with number of RV visits shown in parenthesis. NSS = non\_single\_star flag from Gaia DR3 (0=no binary solution confirmed, 2=spectroscopic binary). RV amp is the rv\_amplitude\_robust \citep{Katz2023}, total amplitude in the RV series after outlier removal (in km/s).
        \item[b] APOGEE DR17 \citep{apogeedr17}. VSCATTER is the scatter in RV (in km/s) between visits, with number of visits in parenthesis. The pipeline reported S/N weighted RV error (VERR) is 0.02-0.06 km/s for all stars in the sample.
        \item[c] Target source. See Table note g in \ref{table:spec_abs}.
        \item[d] Categories are based only on the FIES spectroscopy (Table~\ref{table:spec_abs}).
    \end{tablenotes}
    \end{threeparttable}
\end{table*}
\begin{figure*}
    \centering
    \includegraphics[height=0.9\textheight,width=\textwidth,keepaspectratio]{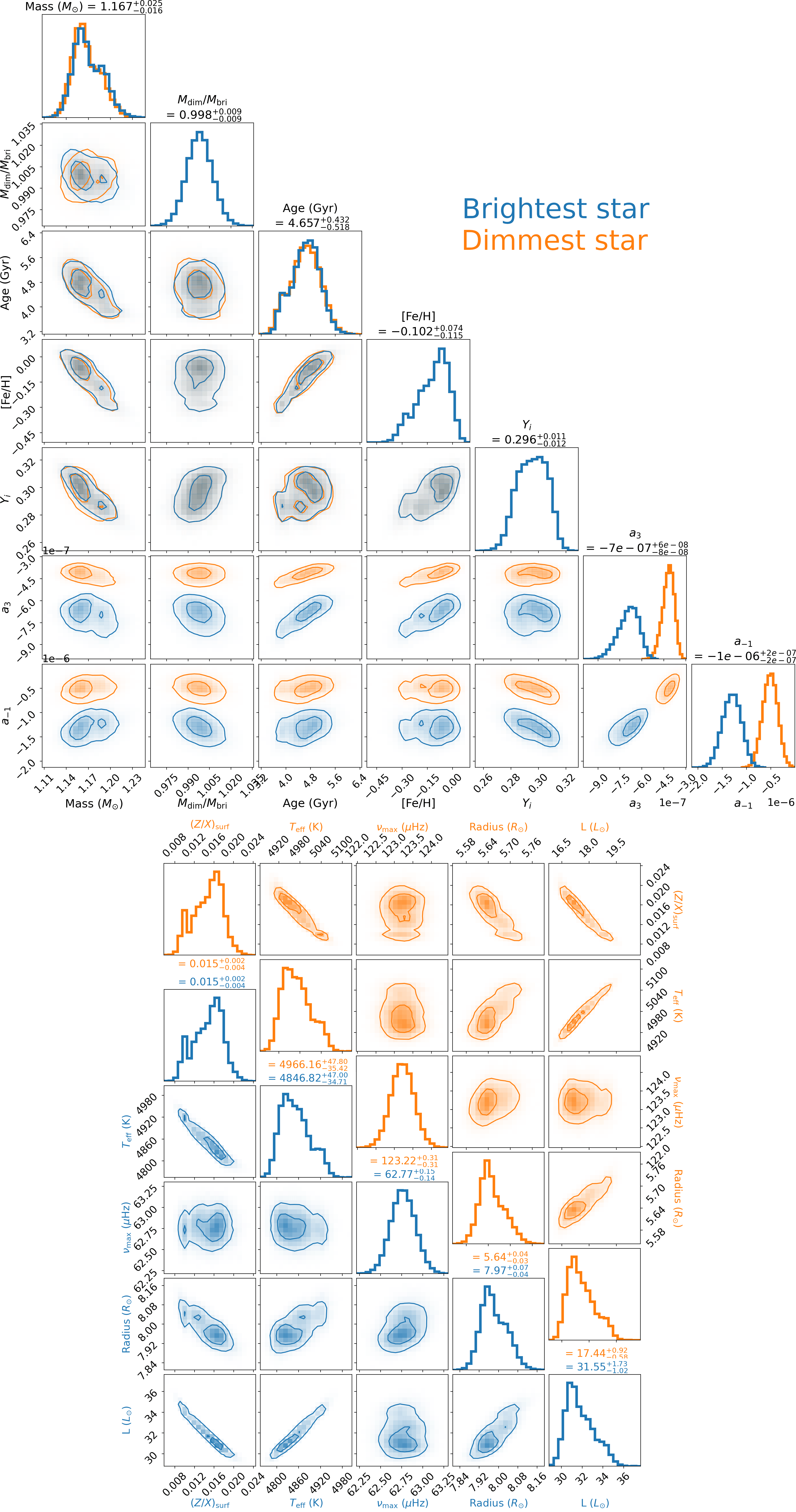}
    \caption[Corner plot of the asteroseismic inference of both stars in KIC\,6501237.]{Corner plot of the asteroseismic inference of both stars in KIC\,6501237. Top: Comparison on the fit parameters, mass, age, metallicity, helium, and surface corrections.  Bottom: Derived parameters for the dimmest (orange) and brightest star (blue). Note that the two corner plots have different axis limits. \label{fig:corner_6501237}}
\end{figure*}
\begin{figure*}
    \centering
    \includegraphics[height=0.9\textheight,width=\textwidth,keepaspectratio]{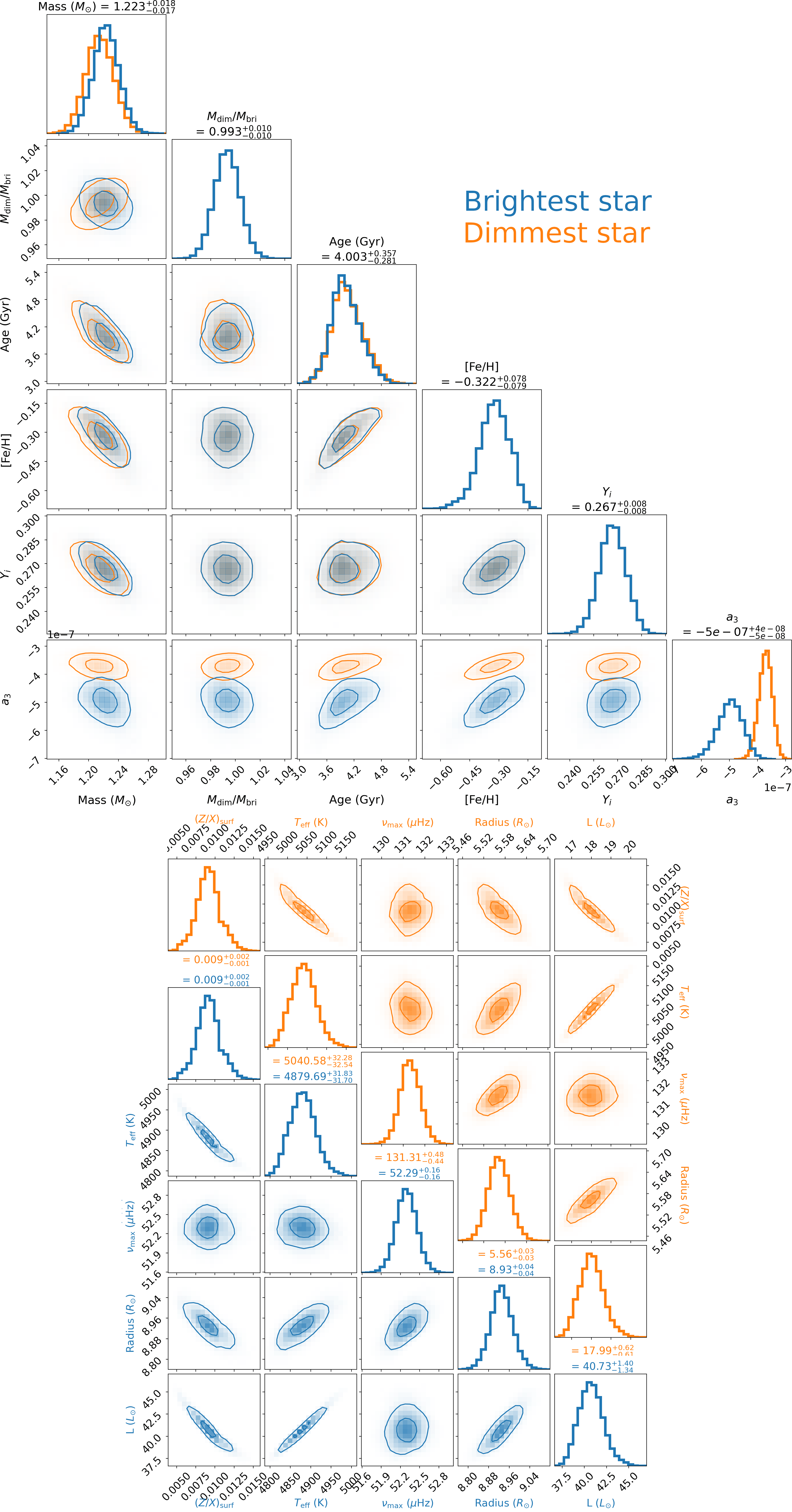}
    \caption[Corner plot of the asteroseismic inference of both stars in KIC\,10592924.]{Corner plot of the asteroseismic inference of both stars in KIC\,10592924. Top: Comparison on the fit parameters, mass, age, metallicity, helium, and surface corrections. Bottom: Derived parameters for the dimmest (orange) and brightest star (blue). Note that the two corner plots have different axis limits. \label{fig:corner_10592924}}
\end{figure*}
\begin{figure*}
    \centering
    \includegraphics[height=0.45\textheight,keepaspectratio]{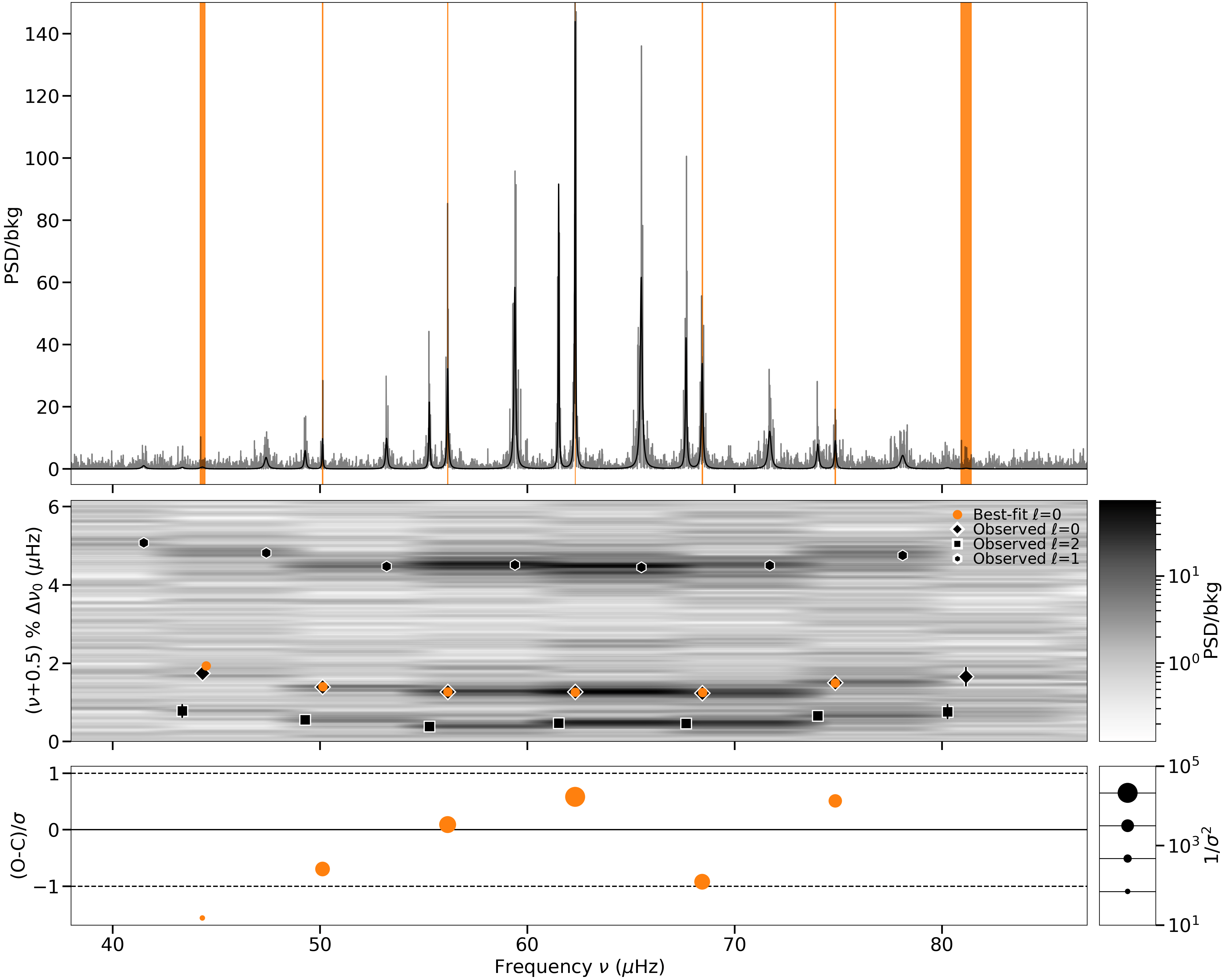}
    \caption[PSD of brightest star in KIC\,6501237, including observed and theoretical frequencies.]{Same as Fig.~\ref{fig:psd_2971380}, but for the brightest star of KIC\,6501237. \label{fig:psd_6501237A}}
\end{figure*}
\begin{figure*}
    \centering
    \includegraphics[height=0.45\textheight,keepaspectratio]{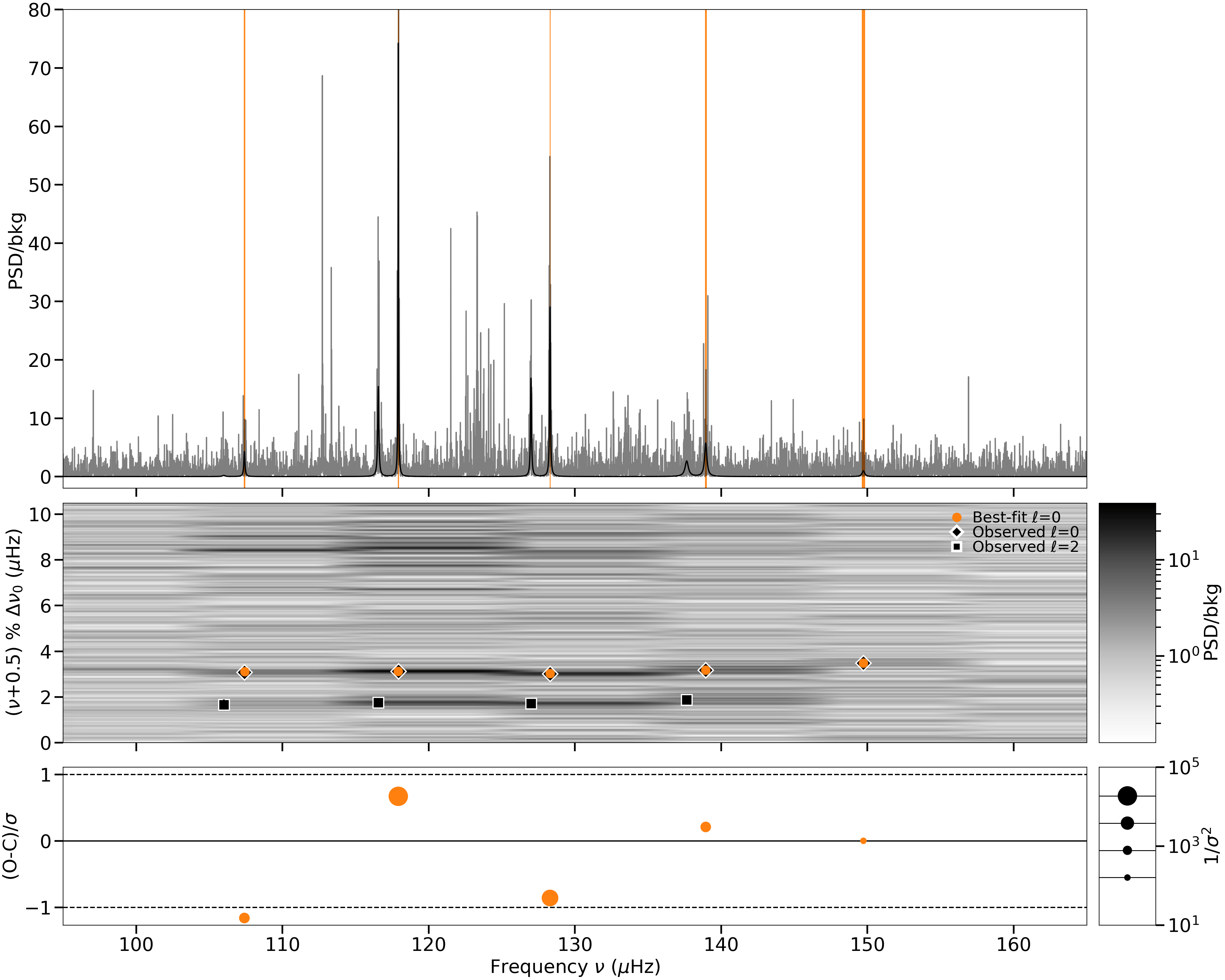}
    \caption[PSD of dimmest star in KIC\,6501237, including observed and theoretical frequencies.]{Same as Fig.~\ref{fig:psd_2971380}, but for the dimmest star of KIC\,6501237. \label{fig:psd_6501237B}}
\end{figure*}
\begin{figure*}
    \centering
    \includegraphics[height=0.45\textheight,keepaspectratio]{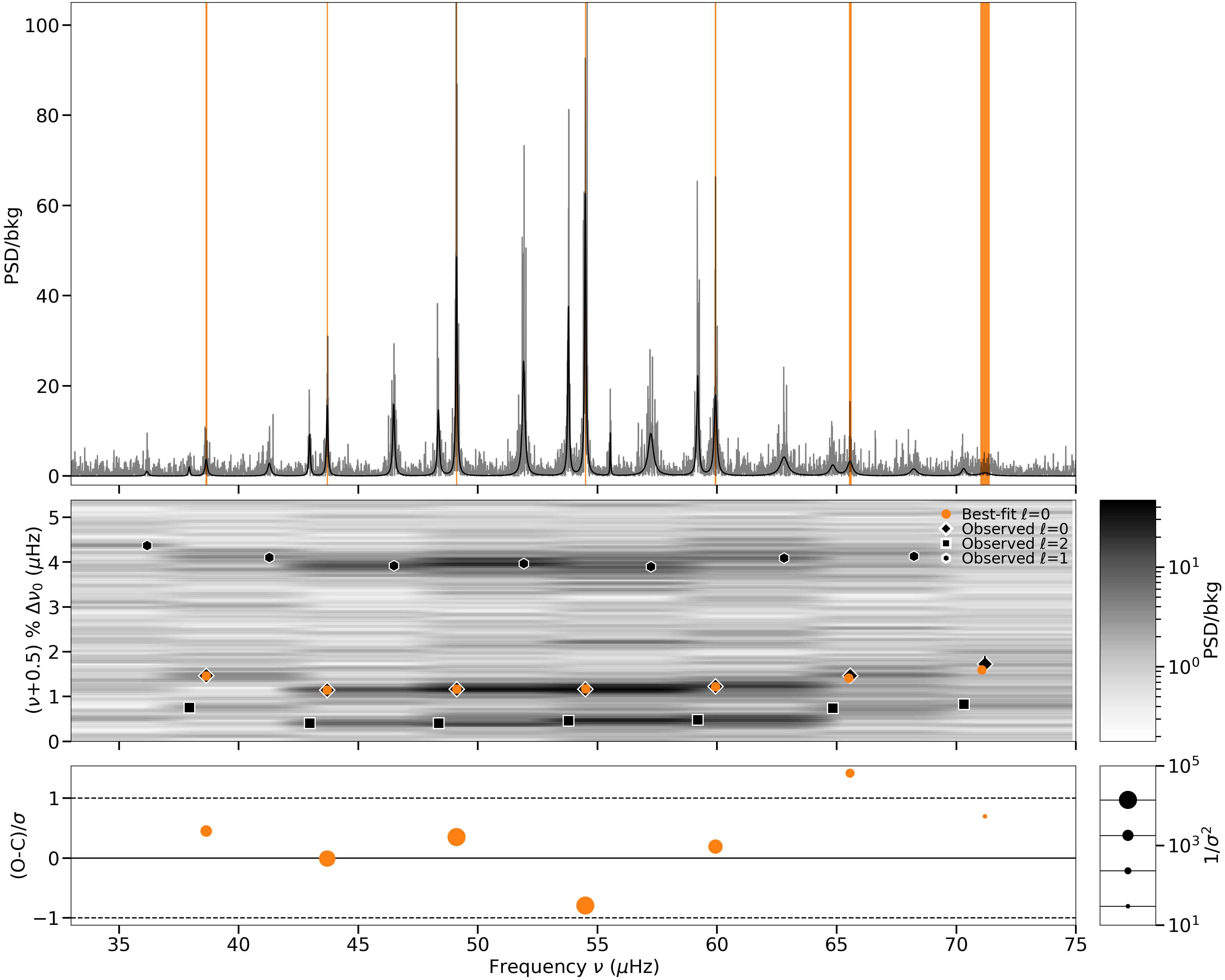}
    \caption[PSD of brightest star in KIC\,10592924, including observed and theoretical frequencies.]{Same as Fig.~\ref{fig:psd_2971380}, but for the brightest star of KIC\,10592924. \label{fig:psd_10592924A}}
\end{figure*}
\begin{figure*}
    \centering
    \includegraphics[height=0.45\textheight,keepaspectratio]{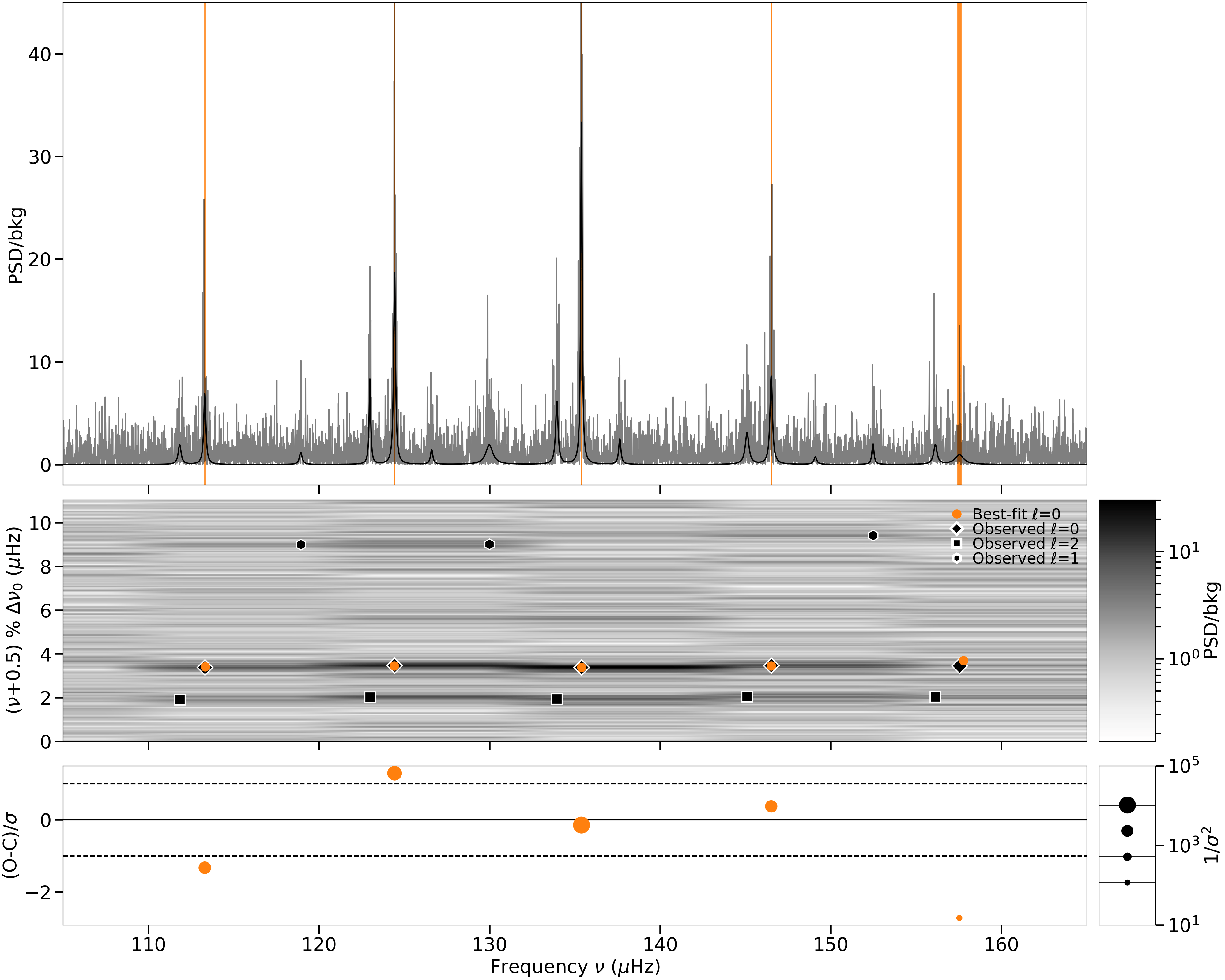}
    \caption[PSD of dimmest star in KIC\,10592924, including observed and theoretical frequencies.]{Same as Fig.~\ref{fig:psd_2971380}, but for the dimmest star of KIC\,10592924. \label{fig:psd_10592924B}}
\end{figure*}

\begin{figure*}
    \centering
    \includegraphics[height=0.45\textheight,keepaspectratio]{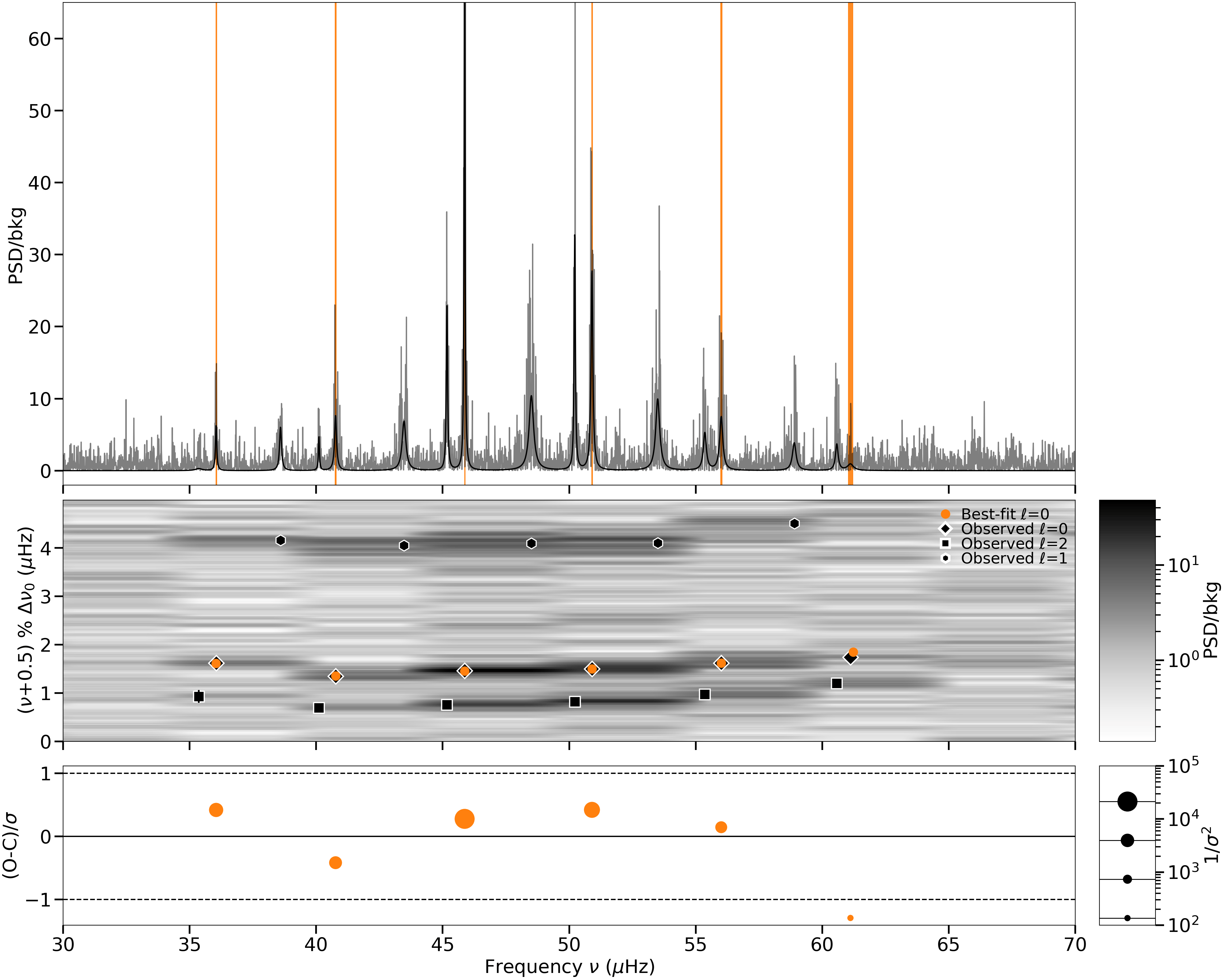}
    \caption[PSD of brightest star in KIC\,11299484, including observed and theoretical frequencies.]{Same as Fig.~\ref{fig:psd_2971380}, but for the brightest star of KIC\,11299484. \label{fig:psd_11299484A}}
\end{figure*}
\begin{figure*}
    \centering
    \includegraphics[height=0.45\textheight,keepaspectratio]{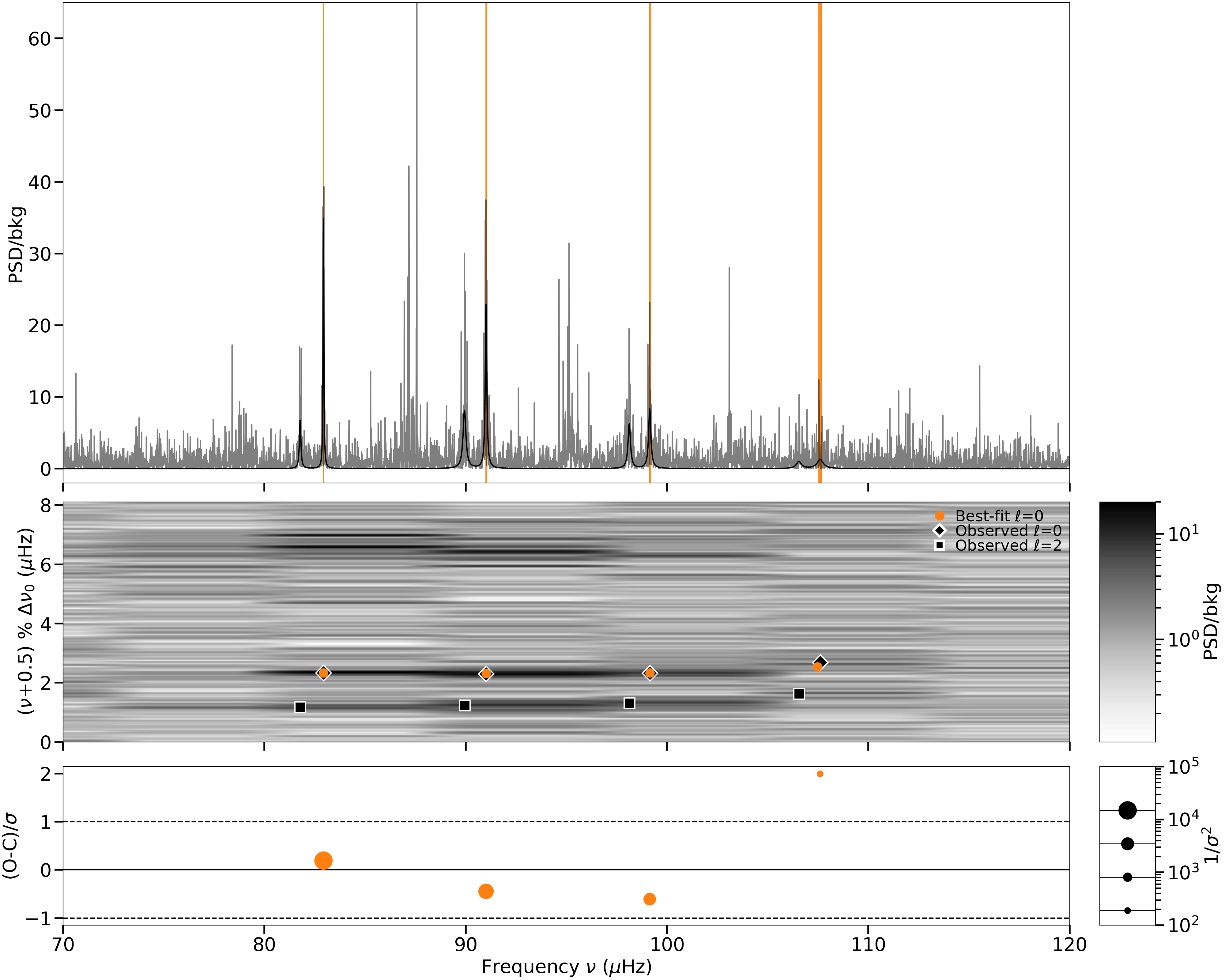}
    \caption[PSD of dimmest star in KIC\,11299484, including observed and theoretical frequencies.]{Same as Fig.~\ref{fig:psd_2971380}, but for the dimmest star of KIC\,11299484. \label{fig:psd_11299484B}}
\end{figure*}
\twocolumn
\begin{table}
\centering
\caption{\label{table:ab_freqs} Peakbagged frequencies for binaries.}
\begin{threeparttable}
\begin{tabular}{lcccc}
\hline
$\ell$       & KIC\,6501237          & KIC\,10592924 & KIC\,11299484         & KIC\,2971380\tnote{b} \\
\hline
1       & 41.497 (52)           & 36.168 (27)   &           &\\
2       & 43.36 (18)           & 37.937 (26)    & 35.36 (13)           &\\
0       & 44.33 (12)           & 38.646 (22)    & 36.047 (14)           & 57.031 (19)\\
1       & 47.398 (27)           & 41.284 (29)   & 38.592 (18)           &\\
2       & 49.284 (17)           & 42.967 (12)   & 40.111 (18)           & 63.372 (48)\\
0       & 50.123 (13)           & 43.704 (10)   & 40.768 (17)           & 64.497 (16)\\
1       & 53.209 (15)           & 46.484 (13)   & 43.473 (24)           &\\
2       & 55.2724 (92)          & 48.353 (13)   & 45.1678 (99)           & 71.2577 (81)\\
0       & 56.1564 (97)          & 49.1114 (84)  & 45.8686 (68)          & 72.4132 (70)\\
3?      &                       &               &        & 73.732 (19)\\
1       & 59.398 (10)           & 51.915 (12)   & 48.505 (21)          &\\
2       & 61.5104 (69)          & 53.7870 (87)  & 50.2217 (92)          & 79.143 (11)\\
0       & 62.3054 (69)          & 54.4933 (85)  & 50.901 (11)          & 80.2257 (90)\\
3?      &                       & 55.5410 (74)  &          &\\
1       & 65.491 (11)           & 57.229 (23)   & 53.499 (21)           &\\
2       & 67.6551 (96)          & 59.189 (12)   & 55.362 (22)           & 87.183 (30)\\
0       & 68.432 (11)           & 59.935 (14)   & 56.011 (21)           & 88.227 (16)\\
3?      &                       &               &         & 89.605 (85)\\
1       & 71.688 (19)           & 62.799 (33)   & 58.896 (33)           &\\
2       & 74.007 (19)           & 64.836 (53)   & 60.578 (23)           & 95.19 (11)\\
0       & 74.851 (16)           & 65.559 (37)   & 61.117 (86)           & 96.355 (30)\\
1       & 78.102 (33)           & 68.223 (69)   &           &\\
2       & 80.26 (19)           & 70.310 (64)    &           &\\
0       & 81.16 (25)           & 71.20 (18)     &            &\\\hline
2       & 105.99 (22)          & 111.832 (42)   & 81.775 (16)       &\\
0       & 107.396 (26)          & 113.306 (18)  & 82.8346 (82)          &\\
1       &                       & 118.929 (51)  &        &\\
2       & 116.552 (14)          & 122.984 (15)  & 89.940 (20)        &\\
0       & 117.9211 (73)         & 124.429 (13)  & 91.010 (12)       &\\
3?      &                       & 126.603 (42)  &        &\\
1       &                       & 129.980 (51)  &        &\\
2       & 126.996 (12)          & 133.943 (21)  & 98.124 (22)        &\\
0       & 128.297 (10)          & 135.3836 (98) & 99.142 (18)       &\\
3?      &                       & 137.630 (28)  &        &\\
2       & 137.645 (46)          & 145.090 (32)  & 106.553 (72)        &\\
0       & 138.945 (26)          & 146.506 (19)  & 107.610 (72)       &\\
3?      &                       & 149.094 (75)          &\\
1       &                       & 152.470 (33)          &\\
2       &                       & 156.126 (42)          &\\
0       & 149.727 (79)          & 157.533 (92)          &\\\hline
\end{tabular}
\begin{tablenotes}
\scriptsize
\item[a] The bar separates the bright and dim stellar components.
\item[b] $\ell$=1 frequencies are reported in Table~\ref{table:29_l1}.
\end{tablenotes}
\end{threeparttable}
\end{table}
\begin{table}
\centering
\caption{\label{table:29_l1} Peakbagged $\ell$=1 frequencies for KIC\,2971380.}
\begin{tabular}{cccc}
\hline
\hline
67.370 (19)     & 75.0051 (61)      & 81.5798 (71)      & 91.686 (23)\\
68.071 (14)     & 75.3706 (60)      & 82.034 (64)       & 92.191 (29)\\
68.3396 (68)    & 75.7982 (45)      & 82.495 (63)       & 92.615 (19)\\
68.8495 (59)    & 76.1408 (54)      & 82.959 (16)       & 93.0945 (73)\\
69.1612 (18)    & 76.4875 (48)      & 83.4818 (43)      & \\
69.4833 (56)    & 76.8045 (63)      & 83.9627 (89)      & \\
69.8227 (96)    & 77.1701 (28)      & 84.3387 (64)      & \\
                & 77.5727 (79)      & 84.6943 (53)      & \\
                & 77.9867 (48)      & 85.1309 (69)      & \\
                &                   & 85.552 (21)       & \\\hline
\end{tabular}
\end{table}
\FloatBarrier
\section{Binary systems of interest}\label{sec:spec_interesting_targets}
\begin{figure}
    \centering
    \includegraphics[width=\columnwidth]{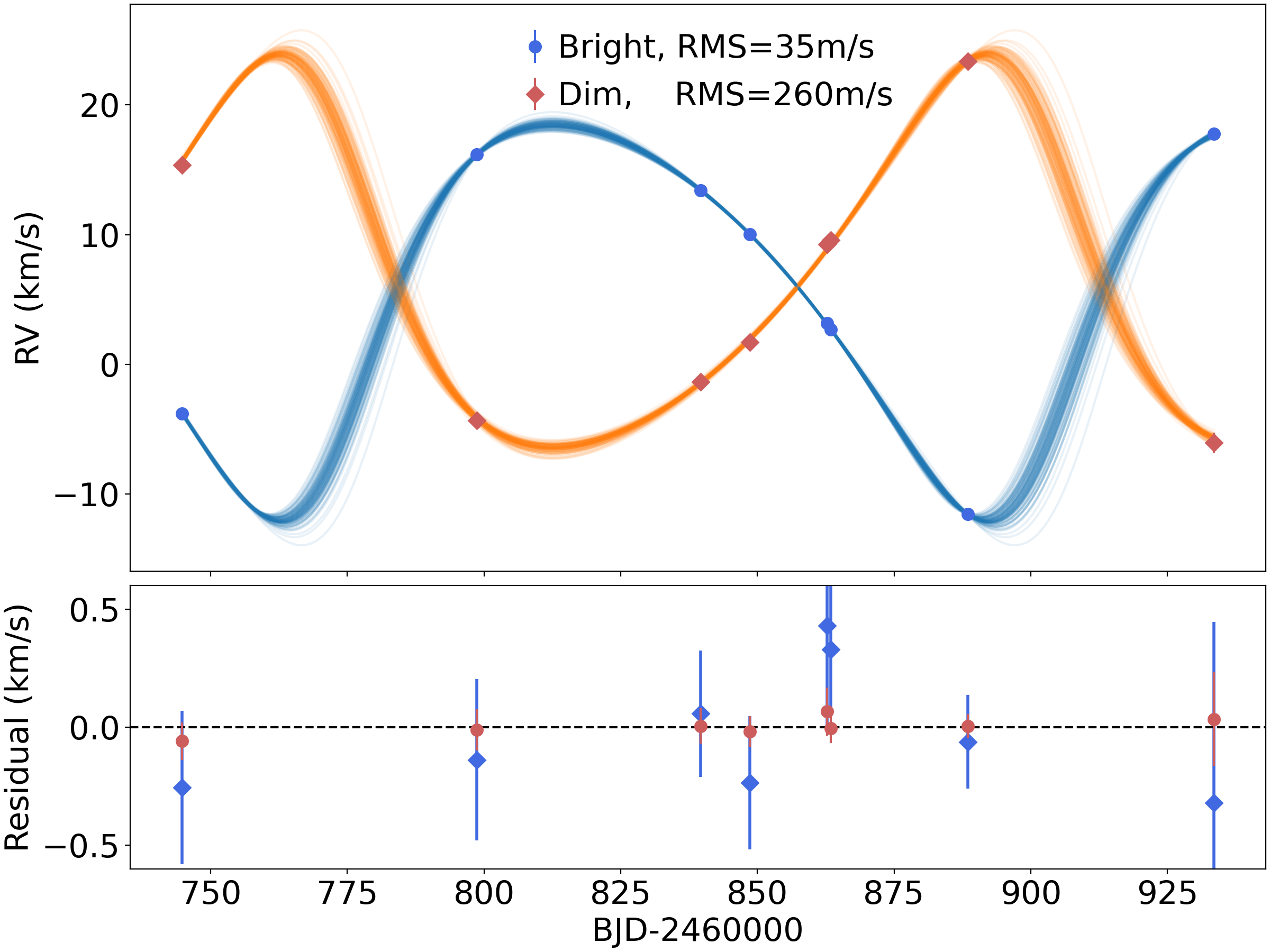}
    \caption{Measured radial velocities for KIC\,2971380, including a fit for the orbital parameters. RVs are offset by +200 km/s.}
    \label{fig:kic2971380_rv}
\end{figure}

In this section we present specific details and analysis for some of the systems where we managed to use the spectroscopic indicators (Sect.~\ref{sec:binary_multi}, \ref{sec:binary_single}, \ref{sec:binary_gaia}) to either confirm, or indicate stars as binaries. We furthermore include a few cases where we disprove binarity, based on the arguments presented in the corresponding sections (KIC\,6369592 \& KIC\,9392647/50). 
\paragraph{KIC\,2971380.}
A low-luminosity RGB star with an observed asteroseismic period spacing of $\Delta P=70.84 \pm 0.03$ \citep{Mosser2018}, and a photometric excess (Mazzi et al., in prep). It has been studied in detail with asteroseismology, as its chemical composition and Galactic kinematics support an ex situ origin, likely from Gaia Enceladus \citep{Montalban2021, Borre2022}. We confirm it as an SB2 system with a luminosity ratio around $25\%$, based on the BF amplitudes.
We performed a Keplerian two-body orbit fit to the RVs and confirm that the mass ratio is consistent with unity at $1\%$ precision, suggesting that the companion is likely evolved past the main sequence as well. The \emph{Kepler} power spectrum does not show indications of oscillations from the dimmer companion. The orbit fit is shown in Fig.~\ref{fig:kic2971380_rv} and Table~\ref{table:kic2971380}, and the RVs are reported in Table~\ref{tab:rv_2971380}.
\begin{table}
\centering
\footnotesize
\caption{\label{table:kic2971380}Spectroscopic orbit properties of KIC\,2971380.}
\begin{threeparttable}
\begin{tabular}{lc}
\hline
\hline
$\rm \frac{Mass_{\rm dim}}{Mass_{\rm bright}}$ & 1.005 (10) \\
$M_{\rm bright} \sin^3 i$ ($\textrm{M}_{\odot}$) & 0.170 (11) \\
$M_{\rm dim} \sin^3 i$ ($\textrm{M}_{\odot}$) & 0.171 (11) \\
$a \sin i$ ($\textrm{R}_{\odot}$) & 75.1 (17) \\
RV semi-amplitude $K_{\rm bright}$ (km/s) & 15.28 (30) \\
RV semi-amplitude $K_{\rm dim}$ (km/s) & 15.20 (34)\\
Eccentricity $e$ & 0.2643 (47) \\
Periastron longitude $\omega$ ($^{\circ}$) & 222.4 (42) \\
Orbital period (days) & 129.26 (33) \\
$t_{0,\rm peri}$ (days)\tnote{a} & 1030.2 (26)\\
RV zero-point $\gamma_{\rm bright}$ (km/s) & -193.82 (22) \\
RV zero-point $\gamma_{\rm dim}$ (km/s) & -194.22 (24)\\
\hline
\end{tabular}
\begin{tablenotes}
\scriptsize
\item[a] Barycentric Julian date minus 2460000.
\end{tablenotes}
\end{threeparttable}
\end{table}
\begin{table}
\centering
    \caption{Radial velocities for KIC\,2971380.}
    \label{tab:rv_2971380}
\begin{threeparttable}
    \begin{tabular}{lccc}
    \hline
BJD-X\tnote{a}   & RV (bright)\tnote{b}  & RV (dim)\tnote{b} & Cov (km$^2$/s$^2$)
\\\hline
744.7479   & -3.778 (79)   & 15.38 (33)   & 0.00595 \\
798.7172   & 16.169 (87)   & -4.30 (34)   & -0.00570 \\
839.6515   & 13.420 (75)   & -1.35 (27)   & 0.00674 \\
848.5864   & 10.030 (66)   & 1.70 (28)   & 0.00888 \\
862.7040   & 3.20 (10)   & 9.25 (45)   & 0.00709 \\
863.4471   & 2.711 (64)   & 9.56 (27)   & 0.00458 \\
888.5140   & -11.547 (52)   & 23.37 (20)   & -0.00002 \\
933.5083   & 17.77 (20)   & -6.03 (77)   & 0.00675\\\hline
\end{tabular}
\begin{tablenotes}
\scriptsize
\item[a] Barycentric Julian date minus 2460000.
\item[b] RV units are in km/s. RVs are offset by +200 km/s.
\end{tablenotes}
\end{threeparttable}
\end{table}
\paragraph{KIC\,4260884, 7697607, 8479383, 10592924.}
These are ABs with separated oscillation envelopes.
We confirm them as double-lined binaries with RV variation. KIC\,8479383 contains at least one core-helium burning (CHeB) star, as evidenced by the star's mixed modes \citep[see e.g.][]{Bedding2011}.
KIC\,7697607 has two oscillators with a small separation in $\nu_{\rm max}$, which means that the oscillation envelopes partially overlap for at least one radial order.

\begin{figure}
    \centering
    \includegraphics[width=0.8\columnwidth]{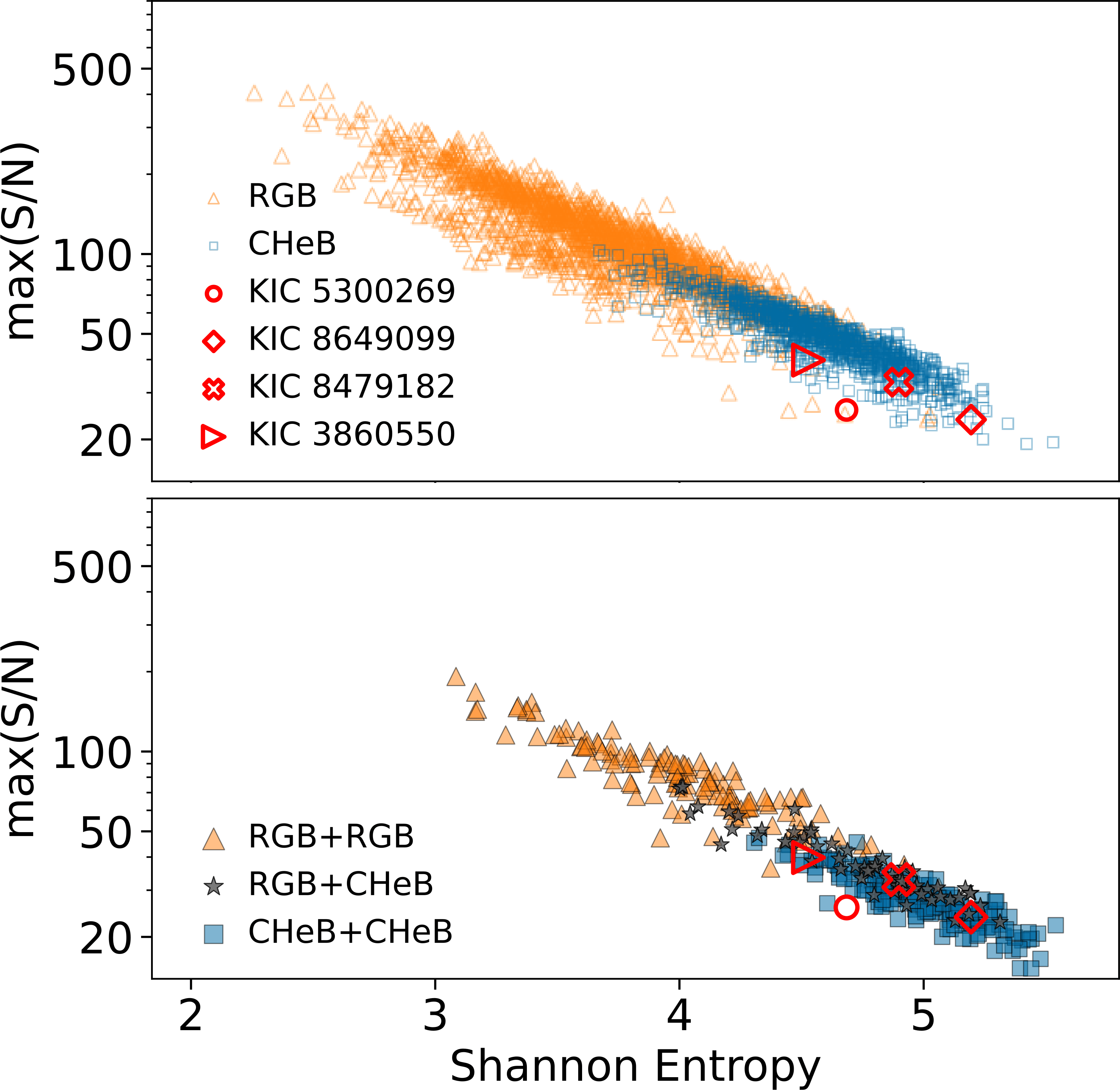}
    \caption{Shannon entropy and maximum S/N of the oscillation envelope, for KIC\,5300269, 8649099, 8479182, and 3860550. Top: Compared with regular stars from \citet{Choi2025}. Bottom: Compared with artificial ABs from \citet{Choi2025}.}
    \label{fig:shannon}
\end{figure}

The power spectrum of KIC\,4260884 appears noisy, particularly for the brightest star.
All its oscillation modes appear complex and broadened, which is unlike a typical CHeB mixed mode pattern \citep[see e.g.][]{Bedding2011}.
\paragraph{KIC\,6501237.}
There are two stars within 3" on-sky, with $\Delta G = 2.3$, with a parallax-derived line-of-sight separation of $0.03\pm 0.6$pc, and Galactocentric velocity difference of only $\sim 2$ km/s. These stars were flagged by \citep{Espinoza2025} as having a bound-binary configuration probability of $50\%$.
We have two spectra of the brightest star, taken $\sim 150$ days apart, showing a small change in its RV and width. Unfortunately, one of the spectra was taken with observational seeing of 2", and likely contains a contribution from both stars, which can explain the RV variation.
Given their close proximity and low RV difference, it is plausible that they are either in a bound orbit or in a close association.
In Sect.~\ref{sec:ab_inference_chap4}, we treat them as a coeval pair, in order to explore if we can obtain a good asteroseismic double-star model fit under this assumption. A seismic confirmation would be interesting as \citet{Miglio2014} predicted that all resolved ABC systems in \emph{Kepler} should be chance alignments. 
\paragraph{KIC\,5300269, 8479182, 8649099, and 3860550.}
These stars have a photometric excess when a seismic estimate of the intrinsic luminosity is compared with the apparent magnitude and Gaia DR3 derived parallax-distance (Mazzi et al, in prep.). They furthermore show complicated oscillation envelopes that could indicate the presence of two oscillation signals overlapping in frequency. We followed \citet{Choi2025}, and measured the Shannon entropy of the oscillation envelope, together with the maximum S/N. This gives a measure of the complexity of the oscillation envelope \citep{Choi2025}. The top panel of Fig.~\ref{fig:shannon} demonstrates that all four systems have comparably high entropy and low S/N relative to the typical \emph{Kepler} field RGs. By comparing them with artificial ABs from \citet{Choi2025}, which were constructed by combining the light curves of two RGs with a flux ratio $\geq 0.9$ and $\nu_{\rm max}$ difference $< 10\%$ (see Fig.~\ref{fig:shannon}, bottom panel), the four stars are compatible with having two overlapping RG oscillators.

KIC\,5300269 shows a change in BF shape over $\sim 120$ days, indicating the presence of two absorption features.
KIC\,8479182 and KIC\,8649099 have photometric luminosities and APOGEE DR17 temperatures that put them at $\sim 2\times L_{\rm RC}$. We confirm them as double-lined and RV variable. Such systems are expected to be the most prevalent \emph{Kepler} ABs \citep{Mazzi2025}.

We only obtained one spectrum for KIC\,3860550. However, the absorption line width and shape indicate the presence of two partially overlapping sets of absorption features of similar strength, which further supports that two stars are present within the aperture. Without a measure of long-term RV variability, we have not confirmed binarity, but it seems unlikely that a chance alignment would have so similar RVs for the two stars.
\paragraph{KIC\,5978324.}
Both Gaia data \cite[see][]{Beck2024} and our spectra confirm KIC\,5978324 as SB1.
The star was studied in \citet{Crawford2024, Crawford2025} as part of a sample of the 48 most massive solar-like oscillators observed by \emph{Kepler}, where they found $M=4.7\pm 0.3 \textrm{M}_{\odot}$.
To date not many oscillating RG binaries in the $4-5 \textrm{M}_{\odot}$ range have been studied in detail, and it is of interest as it is near the mass where stars show Cepheid pulsations during part of their evolution.
\paragraph{KIC\,6369592.}
Both the power spectrum and spectroscopic follow-up show two stars, but the $\nu_{\rm max}$ ratio of $0.516 \pm 0.006$ and BF integral ratio of $\sim 0.1-0.2$ give discrepant indicators for the luminosity ratio. The amplitude ratio of the two asteroseismic oscillation envelopes of $0.218 \pm 0.011$ supports the luminosity ratio estimated from spectroscopy, suggesting that the spectroscopic lines match the two oscillators detected in the power spectrum. The RV separation of $21.2 \pm 0.1$ km/s does not vary over 400 days, which confirms that the two stars cannot be gravitationally bound. The large RUWE value and negative parallax from Gaia could indicate that they are partially resolved by Gaia, despite not having been identified as such.
\paragraph{KIC\,5015647.}
This system shows very high RV variation (19 km/s per year) based on the two spectra taken 141 days apart, which is a clear confirmation of binarity.
Its \emph{Kepler} power spectrum presents a complicated solar-like oscillation pattern at $\sim 30 \mu$Hz, with a forest of narrow modes potentially indicative of an atypical mixed mode coupling strength \citep[see e.g.][]{Takata2016, Matteuzzi2023, vanRossem2024}. In addition, the power spectrum contains a set of higher-amplitude, coherent peaks from $180\mu$Hz until the Nyquist frequency.
The nearby source (8") KIC\,5015657 (Gaia DR3 2053253684738428928) has a magnitude difference of only $\sim 2$ in the Gaia G band. We checked pixel-per-pixel power spectra for the two sources following \citet{Colman2017, Choi2026}, and there is no indication that the nearby source (KIC 5015657) is the cause of either of the two oscillation signals in the power spectrum.
\paragraph{KIC\,4637699, 8295007, 9412408, 10841730, 11299484.}
These show two separated oscillators in their power spectra. From our RVs, some show marginal detections of variability. When combined with Gaia DR3 and APOGEE DR17 indicators of binarity, these all have at least mild indications of spectroscopic and/or astrometric variability from two different sources.
Particularly, KIC\,11299484 clearly presents two absorption features with FIES, and indications of binarity from APOGEE DR17 and Gaia DR3, making it a high-likelihood binary. KIC\,4637699 has an abnormal line shape, Gaia DR3 RUWE=1.1, and mild RV variation with FIES.
\paragraph{KIC\,4282111, 8181509, 11352644, 11465942, 12117790.}
These are not obviously ABCs, but when cross-matched with Gaia astrometry/spectroscopy and APOGEE spectroscopy, they have indications of binarity.
KIC\,8181509, 11352644, and 12117790 have a photometric excess similar to the stars marked in Fig.~\ref{fig:shannon} (Mazzi et al., in prep.). KIC\,12117790 appears to have a second set of oscillations close to the Nyquist frequency, making it an ABC. Additionally, the \emph{Kepler} aperture contains two stars with Gaia G magnitude difference of only $0.6$ and 0.6" optical separation on-sky.
KIC\,11465942 appears visually to have a wide oscillation envelope, and potentially show signs of dipole mode depression \citep[see e.g.][]{Mosser2012, Stello2016}. KIC\,11352644 shows a complex power spectrum, and might contain two overlapping oscillators.
\paragraph{KIC\,9392647 and 9392650.}
These appear to be a chance alignment of two oscillators on the sky, as also found by \citet{Espinoza2025}.
The amplitude ratio of the two oscillation envelopes is different for the two stars' KEPSEISMIC power spectra, which indicates that the ratio depends on which pixels are included in the aperture photometry.
Therefore, the oscillations belong to the two different resolved stars on the sky.

Although KIC\,9392647 and 9392650 have similar distance and sky position, their kinematics rule out a bound configuration. Their RVs differ by $\sim 40 $ km/s, and their proper-motion difference of $\sim 6.6$ mas/yr corresponds to $\sim 62$ km/s. Their angular separation of $6.2 \pm 0.012$" implies a projected separation of at least $\sim 11400$ AU, for which the escape velocity is only $\sim 500$ m/s. This makes the relative velocity far too large for the two resolved stars to be gravitationally bound.

An unresolved companion to one of the stars could in principle shift one measured RV away from the systemic velocity. However, a companion capable of producing an offset of tens of km/s would leave a clear signature, either through an inflated Gaia RUWE or through a larger RV change between our two spectra. We therefore interpret KIC 9392647 and KIC 9392650 as a chance alignment rather than a bound pair. However, we still detect low-significance RV variability for the lines of both stars.
We have not identified the cause of this, but it could be either due to instrumental drift or intrinsic variability \citep{Hekker2008}.
\FloatBarrier
\section{Spectroscopic bias from two stars}\label{sec:spec_binary_bias}
\FloatBarrier
As mentioned in Sect.~\ref{sec:observations}, due to our FIES observations having low S/N, we did not attempt an atmospheric analysis.
However, many of our targets have also been observed at lower spectroscopic resolution ($R\sim 22.500$, $\delta v \sim 13$ km/s) in the near-infrared by APOGEE \citep{Majewski2017}, and have atmospheric parameters available in APOGEE DR17 \citep{apogeedr17}, derived under the assumption that the source only contained a single star.
\begin{figure}
    \centering
    \includegraphics[width=\columnwidth]{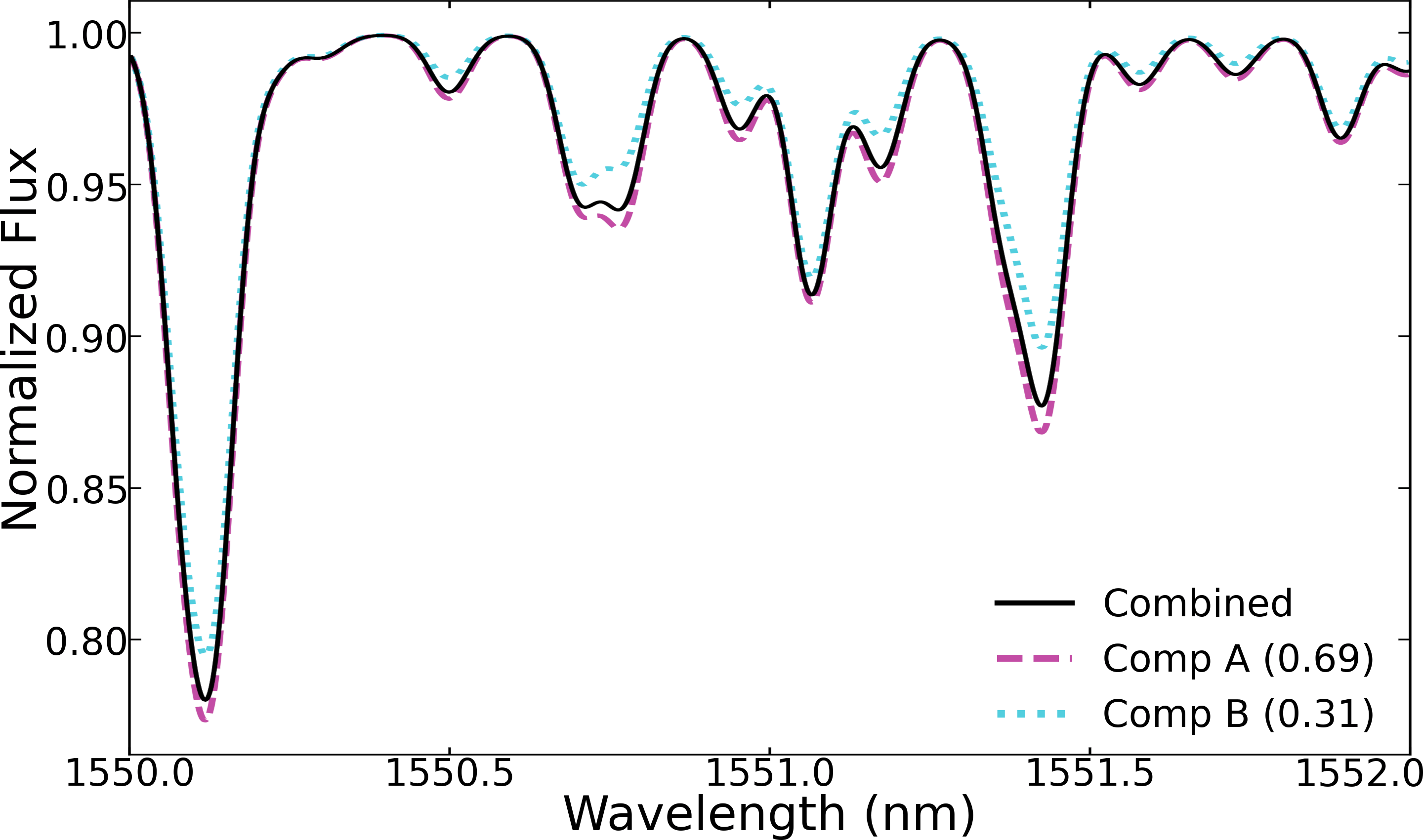}
    \caption{Expected spectroscopic blending of the two stars in KIC\,10592924, in the near-infrared.}
    \label{fig:10592924_spec}
\end{figure}
We evaluate whether this is sufficient to produce an accurate measure of atmospheric parameters for the brightest star, at least within typically adopted uncertainties for effective temperature and metallicity, in a way that is in principle similar to the investigations on unresolved main sequence binaries by \citep{El-Badry2018}. Since the ABs and ABCs that we study have RV separations below the resolution of APOGEE, we can assume that both stars contribute to the lines used for atmospheric analysis. For the optically resolved system KIC\,6501237, it is not certain that the companion contributed to the spectrum, and our investigation is therefore just an upper limit to the expected bias.

We used the metallicity from APOGEE DR17, together with the asteroseismically inferred luminosity ratio, effective temperatures, and $\log g$'s (see Table~\ref{table:ab_inference}). We then constructed synthetic spectra with the same atmospheric parameters, and blended them using the luminosity ratio. 
We computed the synthetic spectral profiles using the 2020 version of the 1D radiative transfer code Turbospectrum \citep{plez2025}, accounting for Non-Local Thermodynamic Equilibrium effects. The synthesis was performed employing standard 1D plane-parallel MARCS model atmospheres \citep{gustafsson2008}. The atomic and molecular line list for the H-band region (1.5-1.7 $\mu$m) was retrieved from the VALD3 database \citep{vald3}. Finally, the synthetic profiles were convolved with a Gaussian profile to match the instrumental resolving power of R=22,000.
For blending the two stars, we neglect the chromatic impact on the light ratio, which is sufficient since the temperature ratio is close to unity. In Fig.~\ref{fig:10592924_spec}, this is illustrated in the near-infrared for KIC\,10592924. We used a Turbospectrum wrapper \citep{gerber2023} to fit the Fe I lines for each spectrum separately (i.e., component A, component B, and the combined spectrum), while keeping the atmospheric stellar parameters fixed. The derived metallicities are consistent within $\pm 0.10$ dex, which is similar to typical observational uncertainties \citep[e.g.][]{Bruntt2010}. With a similar analysis of the temperature, we determined that the brightest star is expected to be within 150K of the observed. Since KIC\,10592924 has the most equal luminosity ratio compared to the difference in effective temperature between the stars, the bias is expected to be smaller for the other stars.
\end{appendix}
\end{document}